\documentclass[a4paper, reqno, 11pt]{amsart}

\usepackage[left=2.5cm, right=2.5cm, top=3cm, bottom=3cm]{geometry}

\usepackage{amsaddr}
\usepackage{amssymb}
\usepackage{amsfonts}
\usepackage{mathrsfs}
\usepackage[T1]{fontenc}
\usepackage[USenglish]{babel}
\usepackage{varioref}
\usepackage{multirow}
\usepackage{array,multicol}
\usepackage{bigdelim}
\usepackage{bigstrut}
\usepackage{bm}
\usepackage{units}
\usepackage{bbold}
\usepackage[font=small, labelfont=bf]{caption}
\usepackage[list=true, labelformat=brace, font=small, labelfont=bf]{subcaption}
\usepackage{graphicx}
\usepackage{csquotes}
\usepackage{longtable}
\usepackage{here}

\usepackage[breaklinks=true
]{hyperref}
\hypersetup{pdftitle={F.Nill: Scaling symmetry and parameter reduction in epidemic SI(R)S models.}}

\usepackage[backend=biber,style=numeric]{biblatex}

\usepackage{enumitem}
\setlist[itemize]{align=parleft,left=0pt..1.5em}
\setlist{noitemsep}

\usepackage{url}

\def\customauthor{\empty}
\def\customdate{\empty}
\let\oldauthor\author
\renewcommand{\author}[1]{\def\customauthor{#1}}
\renewcommand{\date}[1]{\def\customdate{#1}}
\oldauthor{\customauthor\\\customdate}

\theoremstyle{definition}
\newtheorem{definition}{Definition}[section]

\theoremstyle{plain}
\newtheorem{theorem}[definition]{Theorem}
\newtheorem{lemma}[definition]{Lemma}
\newtheorem{corollary}[definition]{Corollary}
\newtheorem{proposition}[definition]{Proposition}

\theoremstyle{remark}
\newtheorem{remark}[definition]{Remark}

\numberwithin{equation}{section}

\newcommand{\ncm}{\newcommand}
\ncm{\rncm}{\renewcommand}

\ncm{\lb}[1]{\label{#1}}

\rncm{\sec}{\setc{0}\section}
\ncm{\bsn}{\bigskip\noindent}
\ncm{\msn}{\medskip\noindent}
\ncm{\ssn}{\smallskip\noindent}
\ncm{\beq}{\begin{equation}}
\ncm{\beqnon}{\begin{equation*}}
\ncm{\eeq}{\end{equation}}
\ncm{\eeqnon}{\end{equation*}}
\ncm{\bea}{\begin{eqnarray}}
\ncm{\beanon}{\begin{eqnarray*}}
\ncm{\eea}{\end{eqnarray}}
\ncm{\eeanon}{\end{eqnarray*}}

\ncm{\ba}{\begin{array}}
\ncm{\ea}{\end{array}}
\ncm{\bpma}{\begin{pmatrix}}
\ncm{\epma}{\end{pmatrix}}
\ncm{\fns}{\footnotesize}

\DeclareMathOperator\sign{sign}
\DeclareMathOperator\const{const.}

\DeclareMathOperator\age{age}

\ncm{\scenA}{\ensuremath{\mathrm{(A)}}}
\ncm{\scenB}{\ensuremath{\mathrm{(B)}}}
\ncm{\scenC}{\ensuremath{\mathrm{(C)}}}
\ncm{\scenD}{\ensuremath{\mathrm{(D)}}}
\ncm{\scenE}{\ensuremath{\mathrm{(E)}}}

\ncm{\scenAp}{\ensuremath{\mathrm{(A^+)}}}
\ncm{\scenBp}{\ensuremath{\mathrm{(B^+)}}}
\ncm{\scenCp}{\ensuremath{\mathrm{(C^+)}}}
\ncm{\scenDp}{\ensuremath{\mathrm{(D^+)}}}
\ncm{\scenEp}{\ensuremath{\mathrm{(E^+)}}}

\ncm{\scenAm}{\ensuremath{\mathrm{(A^-)}}}
\ncm{\scenBm}{\ensuremath{\mathrm{(B^-)}}}
\ncm{\scenCm}{\ensuremath{\mathrm{(C^-)}}}
\ncm{\scenDm}{\ensuremath{\mathrm{(D^-)}}}

\ncm{\scenXpm}{\ensuremath{\mathrm{(X_\pm)}}}

\newcommand{\RR}{\ensuremath{\mathbb{R}}}
\newcommand{\RRN}{\ensuremath{\mathbb{R}_{\geq 0}}}
\ncm{\NN}{\ensuremath{\mathbb{N}}}
\ncm{\ZZ}{\ensuremath{\mathbb{Z}}}
\ncm{\GG}{\ensuremath{\mathbb{G}}}
\rncm{\SS}{\ensuremath{\mathbb{S}}}
\ncm{\bone}{\ensuremath{\mathbb{1}}}
\newcommand{\II}{{ \mathbb{I}}}
\newcommand{\bA}{ \mathbf{A}}
\newcommand{\bB}{ \mathbf{B}}
\newcommand{\bC}{ \mathbf{C}}
\newcommand{\bD}{ \bm{D}}

\newcommand{\bF}{ \mathbf{F}}
\newcommand{\bL}{ \mathbf{L}}

\newcommand{\bM}{ \bm{M}}

\newcommand{\bP}{ \mathbf{P}}
\newcommand{\bp}{ \bm{p}}

\newcommand{\bx}{ \mathbf{x}}

\newcommand{\bLa}{ \bm{\Lambda}}

\newcommand{\bpsi}{ \bm{\psi}}
\newcommand{\bom}{ \bm{\omega}}

\ncm{\pt}{\bp_\tau}
\newcommand{\bfa}{ \mathbf{a}}
\newcommand{\bfb}{ \mathbf{b}}

\newcommand{\bzero}{\mathbf{0}}

\newcommand{\A}{\ensuremath{{\mathcal A}}}
\newcommand{\AH}{\ensuremath{{\mathcal A}^\star}}

\newcommand{\B}{\ensuremath{{\mathcal B}}}

\newcommand{\C}{\ensuremath{{\mathcal C}}}
\newcommand{\D}{\ensuremath{{\mathcal D}}}

\newcommand{\G}{\ensuremath{{\mathcal G}}}
\newcommand{\K}{\ensuremath{{\mathcal K}}}
\newcommand{\N}{\ensuremath{{\mathcal N}}}

\newcommand{\V}{\ensuremath{{\mathcal V}}}
\newcommand{\T}{\ensuremath{{\mathcal T}}}

\rncm{\P}{\ensuremath{{\mathcal P}}}
\ncm{\PPe}{\P_\epsilon}

\rncm{\S}{\ensuremath{{\mathcal S}}}

\renewcommand{\L}{\ensuremath{{\mathcal L}}}
\ncm{\Labc}{\L_{a,b,c}}
\newcommand{\I}{\ensuremath{{\mathcal I}}}

\newcommand{\M}{\ensuremath{{\mathcal M}}}

\rncm{\O}{\mathcal{O}}

\ncm{\Me}{\M_\epsilon}

\ncm{\Pph}{\P_{\mathrm{phys}}}
\ncm{\Tph}{\T_{\mathrm{phys}}}
\ncm{\bTph}{\bar{\T}_{\mathrm{phys}}}
\ncm{\Iend}{\I_{\mathrm{end}}}
\ncm{\Ta}{\T_{\bm{a}}}
\ncm{\Aph}{\A_{\mathrm{phys}}}
\ncm{\Abio}{\A_{\mathrm{bio}}}
\ncm{\Abiop}{\A_{\mathrm{bio},+}}
\ncm{\Asoc}{\A_{\mathrm{soc}}}
\ncm{\Csoc}{\C_{\mathrm{soc}}}
\ncm{\Dsoc}{\D_{\mathrm{soc}}}
\ncm{\Asig}{\A_{\mathrm{split}}}
\ncm{\Csig}{\C_{\mathrm{split}}}
\ncm{\Dsig}{\D_{\mathrm{split}}}
\ncm{\Aone}{\A_{\mathrm{Model-1}}}
\ncm{\Atwo}{\A_{\mathrm{Model-2}}}
\ncm{\Asirs}{\A_{\mathrm{SIRS}}}
\ncm{\Asiso}{\A_{\mathrm{SIS_1}}}
\ncm{\Asist}{\A_{\mathrm{SIS_2}}}
\ncm{\Asisp}{\A_{\mathrm{SIS+}}}
\ncm{\Asism}{\A_{\mathrm{SIS-}}}
\ncm{\Asispm}{\A_{\mathrm{SIS\pm}}}
\ncm{\Dsis}{\D_{\mathrm{SIS}}}
\ncm{\Dsisp}{\D_{\mathrm{SIS+}}}
\ncm{\Dsism}{\D_{\mathrm{SIS-}}}
\ncm{\Dsispm}{\D_{\mathrm{SIS\pm}}}
\ncm{\Asisj}{\A_{\mathrm{SIS_j}}}
\ncm{\Aheth}{\A_{\mathrm{Heth}}}
\ncm{\Dheth}{\D_{\mathrm{Heth}}}
\ncm{\Kheth}{\K_{\mathrm{Heth}}}
\ncm{\Cph}{\C_{\mathrm{phys}}}
\ncm{\Cbio}{\C_{\mathrm{bio}}}
\ncm{\Cone}{\C_{\mathrm{Model-1}}}
\ncm{\Ctwo}{\C_{\mathrm{Model-2}}}
\ncm{\Csirs}{\C_{\mathrm{SIRS}}}
\ncm{\Dph}{\D_{\mathrm{phys}}}
\ncm{\bDph}{\bar{D}_{\mathrm{phys}}}
\ncm{\DBA}{\D_{AB}}
\ncm{\DAB}{\D_{AB}}
\ncm{\Dbio}{\D_{\mathrm{bio}}}
\ncm{\Dbionu}{\D_{{\mathrm{bio},\nu}}}
\ncm{\Dbioz}{\D_{{\mathrm{bio},0}}}
\ncm{\Done}{\D_{\mathrm{Model-1}}}
\ncm{\Dtwo}{\D_{\mathrm{Model-2}}}
\ncm{\DII}{\D_{\RN{2}}}
\ncm{\Dsirs}{\D_{\mathrm{SIRS}}}
\ncm{\Ksirs}{\K_{\mathrm{SIRS}}}
\ncm{\Kbio}{\K_{\mathrm{bio}}}
\ncm{\Kbionu}{\K_{{\mathrm{bio},\nu}}}
\ncm{\Lbio}{\L_{\mathrm{bio}}}
\ncm{\Gdil}{G_{\mathrm{dil}}}
\ncm{\GX}{G_{X}}
\ncm{\GI}{G_{I}}
\ncm{\GS}{G_{S}}

\ncm{\tAph}{\tilde{\A}_{\mathrm{phys}}}
\ncm{\tAbio}{\tilde{\A}_{\mathrm{bio}}}
\ncm{\tAone}{\tilde{\A}_{\mathrm{Model-1}}}
\ncm{\tAtwo}{\tilde{\A}_{\mathrm{Model-2}}}
\ncm{\tAsirs}{\tilde{\A}_{\mathrm{SIRS}}}
\ncm{\TABC}{T(\bA,\bB,\bC)}
\ncm{\Tabc}{\TABC^{\leq 1}}

\ncm{\yph}{y_{\mathrm{phys}}}
\ncm{\Padm}{P_{\mathrm{adm}}}
\ncm{\Plow}{P_{\mathrm{low}}}
\ncm{\Phigh}{P_{\mathrm{high}}}
\ncm{\plow}{p_{\mathrm{low}}}
\ncm{\phigh}{p_{\mathrm{high}}}
\ncm{\Pc}{\P_{\mathrm{cut}}}
\ncm{\Reff}{X_{\mathrm{rep}}}
\ncm{\Reffo}{X_{\mathrm{rep,0}}^*}
\ncm{\Reffe}{X_{\mathrm{rep,end}}^*}
\ncm{\Ie}{I_{\mathrm{end}}^*}
\ncm{\Se}{I_{\mathrm{end}}^*}
\ncm{\dReff}{\dot{X}_{\mathrm{rep}}}
\ncm{\Tosc}{T_{\mathrm{osc}}}
\ncm{\Timm}{T_{\mathrm{imm}}}
\ncm{\Tinf}{T_{\mathrm{inf}}}
\ncm{\Thalf}{T_{\mathrm{half}}}

\ncm{\rvac}{r_{\mathrm{vac}}}

\newcommand{\al}{\alpha}
\newcommand{\la}{\lambda}
\newcommand{\be}{\beta}

\newcommand{\ga}{\gamma}

\newcommand{\Ga}{\Gamma}

\newcommand{\Del}{\Delta}

\newcommand{\ep}{\epsilon}
\newcommand{\vep}{\varepsilon}
\newcommand{\om}{\omega}

\renewcommand{\th}{\theta}

\ncm{\OP}{\Omega_{\PP,\ep}}
\ncm{\oG}{\omega_\G}
\ncm{\ome}{\omega_\ep}
\ncm{\phit}{\varphi_\tau}
\ncm{\p}{\psi}

\ncm{\Aal}{A_\alpha}
\ncm{\Bal}{B_\alpha}
\ncm{\sal}{\sigma_\alpha}

\rncm{\k}{\kappa}
\ncm{\an}{a_\nu}
\newcommand{\re}{{\,\hbox{$\textstyle\triangleright$}\,}}

\newcommand{\reA}{{\,\hbox{$\textstyle\triangleright_\A$}\,}}
\newcommand{\rehA}{{\,\hbox{$\textstyle\triangleright_{\hat{\A}}$}\,}}
\newcommand{\reB}{{\,\hbox{$\textstyle\triangleright_\B$}\,}}

\newcommand{\li}{{\,\hbox{$\textstyle\triangleleft$}\,}}
\newcommand{\up}{{\hbox{\fns $\vartriangle$}}}

\newcommand{\id}{{\rm id}}

\newcommand{\one}{{\mathbf 1}}

\def\cros{\,\raise1.9pt\hbox{$\scriptscriptstyle  > $}\!
          \raise1.5pt\hbox{$\scriptstyle\triangleleft$}\,}
\def\>cros{\cros}
\def\<cros{\,\raise1.5pt\hbox{$\scriptstyle\triangleright$}\!
           \raise1.9pt\hbox{$\scriptscriptstyle < $}\,}
\ncm{\veq}{{\scriptstyle\Vert}}

\ncm{\dR}{\partial_R}
\ncm{\dS}{\partial_S}
\ncm{\dI}{\partial_I}
\ncm{\dD}{\partial_D}
\ncm{\dN}{\partial_N}
\ncm{\dM}{\partial_M}
\ncm{\dX}{\partial_X}
\ncm{\dq}{\partial_q}
\ncm{\dx}{\partial_x}
\ncm{\dy}{\partial_y}
\ncm{\parH}{\partial H}
\ncm{\parHe}{\partial H_{\epsilon}}
\ncm{\parq}{\partial q}
\ncm{\parp}{\partial p}

\ncm{\rto}{\rightarrow}
\ncm{\mto}{\longmapsto}
\ncm{\lto}{\longrightarrow}
\ncm{\Lto}{\Longrightarrow}
\ncm{\lra}{\leftrightarrow}
\ncm{\LRA}{\Leftrightarrow}
\ncm{\LLRA}{\Longleftrightarrow}
\ncm{\LRa}{\Leftrightarrow}
\ncm{\LLRa}{\Longleftrightarrow}

\ncm{\tOP}{\tilde{\Omega}_{\PP,\ep}}
\ncm{\toe}{\tilde{\omega}_\ep}
\ncm{\tHe}{\tilde{H}_{\epsilon}}
\ncm{\tH}{\tilde{H}}
\ncm{\tV}{\tilde{V}}
\ncm{\tK}{\tilde{K}}
\ncm{\tE}{\tilde{E}}
\ncm{\tA}{\tilde{\A}}
\ncm{\tM}{\tilde{\M}}
\ncm{\tI}{\tilde{\I}}
\ncm{\tP}{\tilde{\P}}
\ncm{\tbF}{\tilde{\bF}}
\ncm{\rt}{\tilde{r}_0}
\ncm{\tr}{\tilde{r}_0}
\ncm{\tga}{\tilde{\gamma}}
\ncm{\tal}{\tilde{\alpha}}
\ncm{\tGa}{\tilde{\Gamma}}
\ncm{\Nt}{\tilde{N}}
\ncm{\tHPe}{\tilde{H}_{\PP,\epsilon}}
\ncm{\tre}{\tilde{\rho}_{\epsilon}}
\ncm{\tq}{\tilde{q}}
\ncm{\tp}{\tilde{p}}
\ncm{\ta}{\tilde{a}}
\ncm{\tb}{\tilde{b}}
\ncm{\tbe}{\tilde{\beta}}
\ncm{\tc}{\tilde{c}}
\ncm{\td}{\tilde{d}}
\ncm{\tx}{\tilde{x}}
\ncm{\ty}{\tilde{y}}
\ncm{\tR}{\tilde{R}}
\ncm{\tep}{\tilde{\vep}}
\ncm{\etq}{e^{\tilde{q}}}
\ncm{\etp}{e^{\tilde{p}}}
\ncm{\ttau}{\tilde{\tau}}
\ncm{\trho}{\tilde{\rho}}

\ncm{\hH}{\hat{H}}
\ncm{\hV}{\hat{V}}
\ncm{\hK}{\hat{K}}
\ncm{\hKs}{\hat{K}_\sigma}
\ncm{\hHs}{\hat{H}_\sigma}
\ncm{\hvc}{\hat{v}_C}
\ncm{\hP}{\hat{\P}}
\ncm{\hA}{\hat{\A}}
\ncm{\hAph}{\hat{\A}_{\mathrm{phys}}}
\ncm{\hB}{\hat{\B}}
\ncm{\hC}{\hat{\C}}
\ncm{\hCph}{\hat{\C}_{\mathrm{phys}}}
\ncm{\hD}{\hat{\D}}
\ncm{\hphi}{\hat{\phi}}
\ncm{\crho}{\check{\rho}}
\ncm{\Reflat}{R_1^{\ \flat}}

\ncm{\ok}{\checkmark}
\ncm{\no}{\ding{55}}
\ncm{\s}{\mathsf{s}}
\ncm{\g}{\mathsf{g}}
\ncm{\h}{\mathsf{h}}
\ncm{\HIG}{H_\alpha}
\ncm{\Hal}{H_\alpha}
\ncm{\oIG}{\omega_\alpha}
\ncm{\HPLV}{H_{\mathrm {pLV}}}
\ncm{\omPLV}{\om_{\mathrm {pLV}}}
\ncm{\Hreg}{H^{\mathrm {reg}}}
\ncm{\omreg}{\om^{\mathrm {reg}}}
\ncm{\HPLVreg}{H_{\mathrm {pLV}}^{\mathrm {reg}}}
\ncm{\omPLVreg}{\omega_{\mathrm {pLV}}^{\mathrm{reg}}}
\ncm{\Halreg}{H_{\alpha}^{\mathrm {reg}}}
\ncm{\omalreg}{\omega_{\alpha}^{\mathrm {reg}}}
\ncm{\xnull}{x_{\mathrm {null}}}
\ncm{\Lt}{L_\tau}
\ncm{\Lmin}{L_{\min}}
\ncm{\dLt}{\dot{L}_\tau}
\ncm{\rv}{a_\mathrm{vac}}
\ncm{\dfe}{_\mathrm{dfe}}
\ncm{\en}{_\mathrm{end}}

\ncm{\DBo}{\D_{\Del B>0}}
\ncm{\DelB}{\Del B}

\newcommand{\Eqref}[1]{Eq. \eqref{#1}}

\ncm{\ulim}[1]{\underset{#1}{\lim}}
\ncm{\secref}[1]{Section \ref{#1}}
\ncm{\figref}[1]{Fig. \ref{#1}}

\newcommand{\minus}{\scalebox{0.75}[1.0]{$-$}}
\newcommand{\inv}{^{\minus 1}}
\ncm{\vsir}{V_{SIR}}
\ncm{\Vsir}{V_{SIR}}
\ncm{\hsir}{H_{SIR}}

\ncm{\zit}[1]{\autocite{#1}}
\ncm{\GHS}{\textsc{HGS }}
\ncm{\Upot}{$U\!$-potential}
\ncm{\QUpot}{quasi-\Upot}
\ncm{\UE}{{V^E}}
\ncm{\VE}{{V^E}}
\ncm{\alpm}{\upsilon_\pm}
\ncm{\alp}{\upsilon_+}
\ncm{\alm}{\upsilon_-}
\ncm{\vpm}{\upsilon_\pm}
\ncm{\xpm}{x_\pm}
\ncm{\qpm}{q_\pm}

\ncm{\vp}{\upsilon_+}
\ncm{\vm}{\upsilon_-}
\ncm{\vO}{v_{\O}}
\ncm{\vN}{v_N}
\ncm{\vt}{v_\tau}
\ncm{\ut}{u_\tau}
\ncm{\apm}{a_\pm}
\ncm{\bpm}{b_\pm}
\ncm{\upm}{u_\pm}
\ncm{\qp}{q_+}
\ncm{\qc}{q_c}
\ncm{\upl}{u_+}
\ncm{\xp}{x_+}
\ncm{\xc}{x_c}
\ncm{\epm}{\varepsilon_\pm}
\ncm{\fpm}{f_\pm}
\ncm{\Apm}{A_\pm}
\ncm{\Bpm}{B_\pm}
\ncm{\Dpm}{D_\pm}
\ncm{\Dp}{\vp}
\ncm{\Dc}{\Delta_c}

\ncm{\Spm}{S_\pm}
\ncm{\rSpm}{\rho S_\pm}
\ncm{\thpm}{\theta_\pm}

\ncm{\ntg}{\notag\\}
\ncm{\Ss}{S_{\textsl{sample}}}
\ncm{\Is}{I_{\textsl{sample}}}
\ncm{\Zs}{Z_{\textsl{sample}}}
\ncm{\Es}{E_{\textsl{sample}}}
\ncm{\Ns}{N_{\textsl{sample}}}
\ncm{\rhos}{\rho_{\textsl{sample}}}
\ncm{\gs}{\gamma_{\textsl{sample}}}
\ncm{\Zrge}{Z_{\rho,\gamma,E}}
\ncm{\Zmax}{Z_{\max}}
\ncm{\Qmax}{Q_{\max}}
\ncm{\el}{e^{\lambda}}
\ncm{\Ve}{V_{\epsilon}}
\ncm{\He}{H_{\epsilon}}

\ncm{\HPe}{H_{\PP,\epsilon}}
\ncm{\Emax}{E_{\max}}
\ncm{\rep}{\rho_{\epsilon}}
\ncm{\qe}{q_{\epsilon}}
\ncm{\expe}{\exp_{\epsilon}}
\ncm{\lne}{\ln_{\epsilon}}
\ncm{\Vpme}{V_{\pm,\ep}}
\ncm{\Vpe}{V_{+,\ep}}
\ncm{\Vme}{V_{-,\ep}}
\ncm{\qpme}{q_{\pm,\ep}}
\ncm{\qpe}{q_{+,\ep}}
\ncm{\qme}{q_{-,\ep}}
\ncm{\xpme}{x_{\pm,\ep}}
\ncm{\xpe}{x_{+,\ep}}
\ncm{\xme}{x_{-,\ep}}
\ncm{\qG}{q_\G}
\ncm{\pG}{p_\G}
\ncm{\ys}{y_2^*}
\ncm{\xs}{x_1^*}
\ncm{\vs}{v_2^*}
\ncm{\us}{u_1^*}
\ncm{\fl}{\varphi_\tau}
\ncm{\Gas}{\Gamma_\sigma}
\ncm{\tmp}{\age}
\ncm{\tImp}{\tau_{\I,\max}}

\begin{document}

\title[Symmetry and parameter reduction in  SI(R)S models]
{Scaling symmetries and parameter reduction\\ in epidemic SI(R)S models.}
\author{Florian Nill}

\address{Department of Physics, Free University Berlin, Arnimallee 14, 14195 Berlin, Germany.}
\date{\today}
%
\email{florian.nill@fu-berlin.de}



\begin{abstract}
Symmetry concepts in parametrized dynamical systems may reduce the number of external parameters by a suitable normalization prescription. If, under the action of a symmetry group $\G$, parameter space $\A$ becomes a (locally) trivial principal bundle,
$\A\cong\/A/\G\times\G$, then the normalized dynamics only depends on the quotient $\A/\G$. In this way, the dynamics of fractional variables in 
homogeneous epidemic SI(R)S models, with standard incidence, absence of $R$-susceptibility and compartment independent birth and death rates, turns out to be isomorphic to (a marginally extended version of) Hethcote's classic endemic model, first presented in 1973.
The paper studies a 10-parameter master model with constant and $I$-linear vaccination rates, vertical transmission and a vaccination rate for susceptible newborns. As recently shown by the author, all demographic parameters are redundant. After adjusting time scale, the remaining 5-parameter model admits a 3-dimensional abelian scaling symmetry. By normalization we end up with Hethcote's extended 2-parameter model. Thus, in view of symmetry concepts, reproving theorems on endemic bifurcation and stability in such models becomes needless.
\end{abstract}

\maketitle



\section{Introduction}
\begin{figure}[H]
\includegraphics[width=0.5\textwidth]{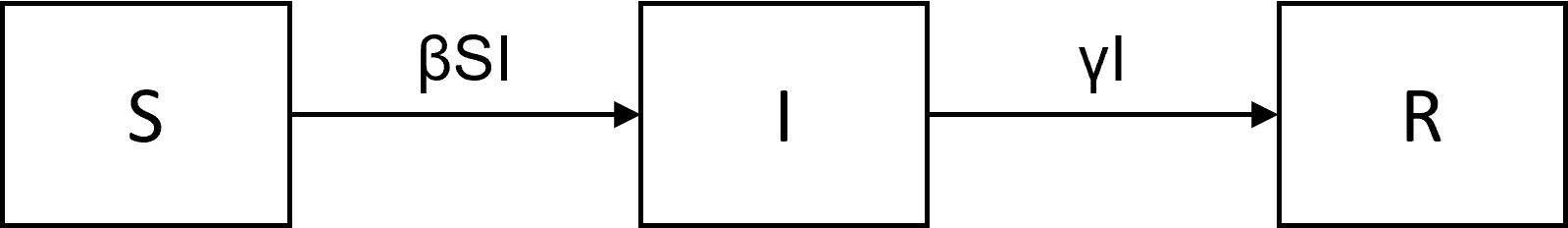}
	\caption{Flow diagram of the SIR model.}
	\label{Fig_SIR}
\end{figure}

The classic SIR model was introduced by Kermack and McKendrick in 1927~\cite{KerMcKen} as one of the first models in mathematical epidemiology. The model divides a population into three compartments with fractional sizes $S$ (Susceptibles), $I$ (Infectious) and $R$ (Recovered), such that $S+I+R=1$. The flow diagram between compartments, as given in Figure \ref{Fig_SIR}, leads to the dynamical {system}
\begin{equation}
\dot{S}=-\be SI,\qquad\dot{I}=\be SI-\ga I,\qquad\dot{R}=\ga I.
\label{SIR}
\end{equation}

\vspace{-9pt}

Here, $\ga$ denotes the recovery 
rate and $\be$ the effective contact rate (i.e., the number of contacts/time leading to infection of a susceptible, given the contacted was infectious). Members of $R$ are supposed to be immune forever. Due to \eqref{SIR}, $S$ decreases monotonically, eventually  causing  
$\be S<\ga$ and $\dot{I}<0$. At the end, the disease dies out, $I_\infty=0$, and one stays with a nonzero final size 
$S_\infty>0$, thus providing a model for {\em Herd immunity}. 

\begin{figure}[H]

\hspace{2.2cm}
\begin{subfigure}[H]{0.4\linewidth}
\includegraphics[width=\textwidth]{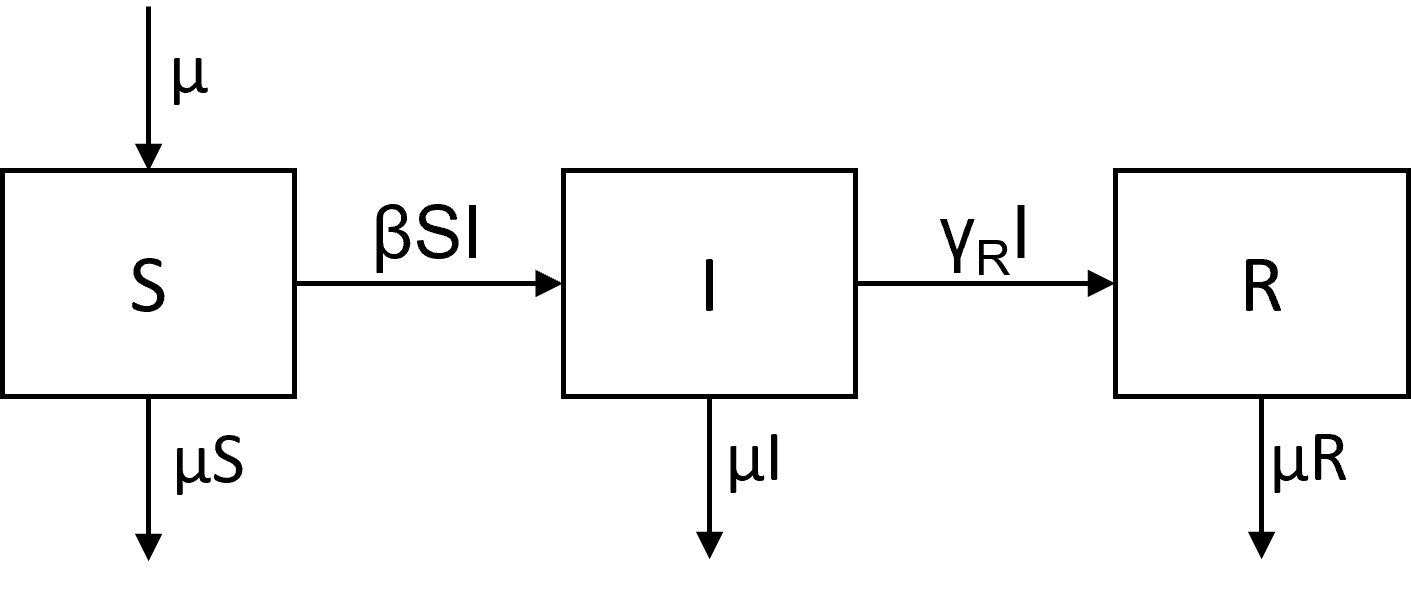}
	\caption{\centering Heathcote's model}
\end{subfigure}

\begin{subfigure}[H]{0.4\linewidth}
\includegraphics[width=\textwidth]{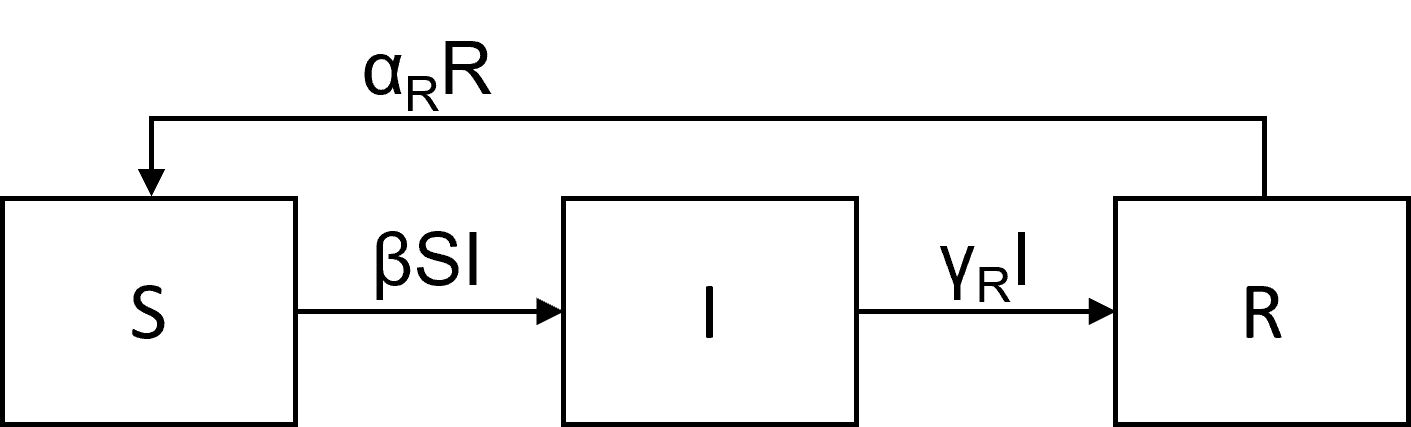}
	\caption{\centering SIRS model}
\end{subfigure}
\qquad
\begin{subfigure}[H]{0.25\linewidth}
\includegraphics[width=\textwidth]{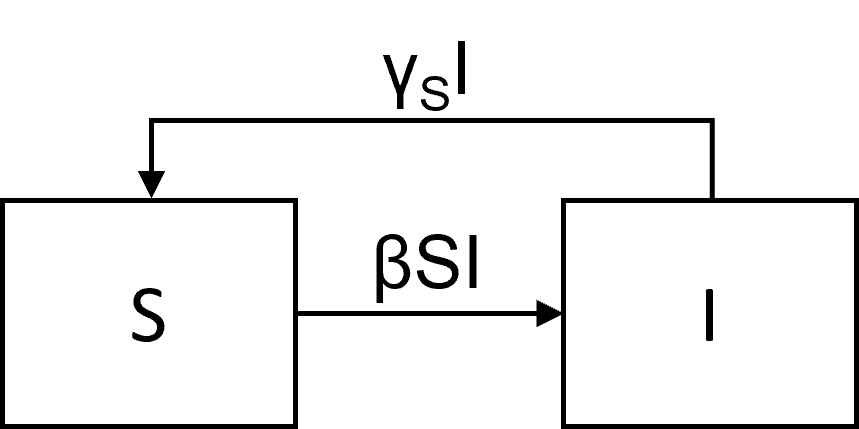}
	\caption{\centering SIS model}
\end{subfigure}
\caption{Standard models featuring {endemic equilibria.}}
	\label{Fig_endemic}
\end{figure}

To construct models also featuring {\em endemic}, scenarios one needs enough supply of susceptibles to keep the incidence $\be SI$ ongoing above a positive threshold. The literature discusses three basic methods to achieve this, see Figure \ref{Fig_endemic}.

\begin{itemize}
\item
Heathcote's {\em {classic endemic model}}
adds balanced birth and death rates $\mu$ to the SIR model and assumes all newborns are susceptible. This leads to a 
bifurcation from a stable disease-free equilibrium point to a stable endemic
scenario when raising the basic reproduction number 
$r_0=\be/(\ga_R+\mu)$ above one~\cite{Hethcote1974, Hethcote1976, Hethcote1989, Hethcote2000}.
\item
The {\em SIRS model} adds an immunity waning flow,
$\al_R R$ from $R$ to $S$, to the SIR model, leading to the same result, with $r_0=\be/\ga_R$
\item
The {\em SIS model} considers recovery without immunity, i.e., a recovery flow $\ga_S I$ from $I$ to $S$, while putting $R=0$. Again, this leads to the same result, with $r_0=\be/\ga_S$.
\end{itemize}

\vspace{-9pt}

Since the original work by Hethcote the literature on endemic bifurcation in SI(R)S-type models is vast. For a comprehensive and self-contained overview of the history, methods and results in mathematical epidemiology see the textbook by M. Martcheva~\cite{Martcheva}, wherein an extensive list of references to original papers is also given. One should also mention that SIR-type models always neglect incubation times, so in this paper I will not look at SEIR models taking care of that.

\bsn
This paper is based on the idea that, when adding more parameters to standard models, reproving theorems may become superfluous, if instead there is a symmetry operation ``turning parameter space north'', i.e. mapping  the seemingly more complicated model to the simpler one. Obvious examples would be diagonalizing the matrix $\bA$ in a linear system $\dot{\bx}=\bA\bx$ or rotating a constant external (say magnetic) force field 
$\bB=(B_1,B_2,B_3)$ in a system of radially interacting particles such that $g\bB=(0,0,|\bB|),\ g\in SO(3)$.
In SI(R)S-type models the simplest example has been proposed in  \cite{Nill_Redundancy}, showing that,  under quite general conditions, demographic  parameters are explicitly redundant when looking at fractional variables (so, this may be viewed as a translation symmetry in parameter space). Applying this, for example, to a recent paper on backward bifurcation in a variable population SIRS model with $R$-susceptibility by \cite{AvramAdenane2022}, many results of that paper  already follow from earlier results in~\cite{KribsVel} and~\cite{LiMa2002}.

\bsn
Going one step ahead, in this paper I will analyze parameter symmetries in a homogeneous 10-parameter SI(R)S $\equiv$ combined SIRS/SIS model with standard incidence, four demographic parameters and a continuous vaccination flow from $S$ to $R$.  Motivated by common observations in the Covid-19 pandemic, I will additionally consider a vaccination rate being proportional to $I$. This models diminishing willingness to get vaccinated when public data indicate decreasing prevalence. 
Using redundancy of birth and death rates and three scaling transformations, we will see that such an elaborated model in fact still boils down to a marginally extended version of Hethcotes classic endemic model. This generalizes earlier results in~\cite{Nill_Omicron}. As a particular consequence, an $I$-linear vaccination rate may always be transformed to zero. In summary, 
by symmetry arguments reproving theorems on endemic bifurcation and stability in such models becomes needless.

\bsn
The plan of this paper is as follows. Section \ref{Sec_Heth} gives a self contained review on Hethcote's endemic model, thereby introducing basic terminology and notation. To prepare the setting for symmetry operations, I will look at a mathematically slightly extended version, by allowing also negative values of $S$ and possibly negative recovery rates $\ga_R>-\mu$. Standard techniques for proving endemic bifurcation and stability immediately generalize to this setting. Identifying $\ga:=\ga_R+\mu>0$ as a pure time scale, this model essentially depends on just two parameters.

Section \ref{Sec_SIRS-model} introduces the 10-parameter SI(R)S model.
After removing redundant demographic parameters and performing two more transformation steps, we will see that for a wide range of parameters $\bfa\in\AH$, including all epidemiologically interesting scenarios, this model actually becomes isomorphic to the extended 2-parameter Hethcote model. In absence of immunity waning and constant vaccination, but still with an $I$-linear vaccination and two recovery flows from $I$ to $R$ and $S$, respectively, the model even becomes isomorphic to the classic SIR model, see Subsection 
\ref{Sec_quasi-SIR}.

Section \ref{Sec_symmetries} provides a group theoretical approach to explain this scenario. There is a coordinate free concept of parameter symmetry as a group $\G$ acting on phase$\times$parameter space,
$\P\times\A$, projecting to an action of $\G$ on $\A$, and leaving the dynamical system form invariant. In our SI(R)S model, choosing suitable coordinates in $\P\times\A$, the group $\G$ is easily identified as a composition of scaling transformations of, respectively, $S,I$ and $x-1$, $x$ being the replacement number. The action of 
$\G$ on $\A$ leaves the above sub-range $\AH\subset\A$ invariant and turns $\A$ into a principal $\G$-bundle. Moreover, $\AH\cong\AH/\G\times\G$ as trivial principal bundles, and any such trivialization induces an isomorphism mapping the original  system with parameter space $\AH$ to an equivalent system with reduced parameter space $\AH/\G$. In this way, the 2-parameter space of the extended Hethcote model is identified as the quotient $\AH/\G$. We also have a ``gauge fixing result'' result, showing that any parameter configuration $\bfa\in\AH$ is 
$\G$-equivalent to a configuration 
$\bfa'\in\Abio$, where $\Abio\subset\A$ denotes the subset of epidemiologically admissible parameters.

Appendix \ref{App_KorW} relates the Korobeinikov-Wake SIRS model in
\cite{KorobWake} to the present formulation in Section \ref{Sec_SIRS-model}, showing that without additional parameter constraints their model may possibly lead to non-physical equilibrium states, $S^*>1$.
Appendix \ref{App_alpha=0} provides a Hamiltonian approach to the ''quasi-SIR case" in Section \ref{Sec_quasi-SIR} and Appendix \ref{App_SIS+vacc} shortly describes the exceptional case of a SIS model with $I$-linear vaccination not covered within the main setting.

\section{Hethcote's endemic model revisited\label{Sec_Heth}}
In this Section we introduce notation and terminology and shortly review Hethcote's classic results~\cite{Hethcote1974, Hethcote1976, Hethcote1989, Hethcote2000}. Replacing $R=1-S-I$, Hethcote's model in Fig. \ref{Fig_endemic} leads to the dynamical system

$$
\dot{S}=-\be SI +\mu(1-S),\qquad\dot{I}=\be SI-(\ga_R+\mu)I
$$
Introducing dimensionless variables 
\begin{equation}
\begin{aligned}
\ga	&:=\ga_R+\mu\qquad	& r_0&:=\be/\ga	&\qquad
a&:=\mu/\ga
\\
x	&:=r_0 S	& y&:=r_0 I	& \tau&:=\ga t
\end{aligned}
\label{xy}
\end{equation}
we get
\begin{equation}
\begin{aligned}
\ga\inv\dot{x}	&=-xy + a(r_0-x)\,,
\\
\ga\inv\dot{y} &=(x-1)y\,.
\end{aligned}
\label{dot_xy}
\end{equation}
Here, $r_0$ is the {\em basic reproduction number} (also called {\em contact number} by Hethcote), i.e. the average number of new cases produced by one infected in a totally susceptible population, and  $x$ is the {\em effective reproduction number} (also called {\em replacement number} by Hethcote), i.e. the average number of new cases produced by one infected at time $t$. More generally, the replacement number $x$ could be defined as the ratio inflow/outflow at the $I$-compartment, making the second equation in \eqref{dot_xy} universal by definition. In particular, endemic equilibria always satisfy $x=1$. Looking at domains of definition, Eqs. \eqref{xy} imply
$$
(x,y)\in \RRN^2\,,\qquad x+y\leq r_0\,,\qquad
0\leq a	<1\,,\qquad r_0>0\,.
$$
Also, $\ga>0$ only sets time scale, i.e. without loss one could choose $\ga=1$ and use $\tau$ as a rescaled time variable. We now slightly extend the above restrictions by using the 
\begin{definition}\label{Def_Heth}
By the {\em extended Hethcote model} we mean the dynamical system \eqref{dot_xy} on phase space $\P=\{(x,y)\in\RR\times\RRN\}$ with parameters $\ga>0,\ a>0,\ r_0\in\RR$.
\end{definition}
\noindent
Hethcote's original methods immediately apply to this extended version. First note
\begin{lemma}\label{Lem_Heth}
For any initial condition $\bp_0\in\P$ the forward flow of the system \eqref{dot_xy} stays bounded.
\end{lemma}
\begin{proof}
Let $x_-\leq\min\{0,r_0\}$ and $x_+\geq\max\{0,r_0,ar_0+(1-a)x_-\}$ and denote $T(x_-,x_+)\subset\P$ the triangle with corners 
$\bp_\li=(x_-,0)$, $\bp_\re=(x_+,0)$ and $\bp_{\up}=(x_-,x_+-x_-)$. Since given $\bp_0\in\P$ we can always choose $x_\pm$ as above such that 
$\bp_0\in T(x_-,x_+)$, it is enough to show that $T(x_-,x_+)$ is forward invariant. Since obviously $y=0\Rightarrow\dot{y}=0$ and $x=x_-\Rightarrow\dot{x}\geq 0$ we are left to show that on the diagonal
$0\leq y=x_+-x\leq x_+-x_-$ we have $\dot{x}+\dot{y}\leq 0$. Assuming $a\leq 1$ we get
$\dot{x}+\dot{y}=a(r_0-x_+)+(a-1)y\leq 0$. If instead 
$a>1$ then  $\dot{x}+\dot{y}\leq a(r_0-x_+)+(a-1)(x_+-x_-)\leq 0$.
\end{proof}

Next, using $y\inv$ as a Dulac function as in \cite{Hethcote1976, Hethcote1989}, one immediately checks 
$\partial_x(\dot{x}/y)+\partial_y(\dot{y}/y)=-(1+a/y)<0$, and so, by the Bendixson–Dulac theorem, in $\P$ there exist no periodic solutions, homoclinic loops or oriented phase polygons of the dynamical system \eqref{dot_xy}. Thus we arrive at

\begin{theorem}[Hethcote 1973~\cite{Hethcote1974}]\label{Thm_Heth}
For any initial conditon $\bp_0=(x_0,y_0)\in\P$ the forward orbit 
$\phi_t(\bp_0)$ of the dynamical system \eqref{dot_xy} exists for all $t>0$. If $r_0\leq 1$ or $y_0=0$, then $\lim_{t\rto\infty}\phi_t(\bp_0)=(r_0,0)$. Otherwise 
$\lim_{t\rto\infty}\phi_t(\bp_0)=(1,a(r_0-1))$.
\end{theorem}

\begin{proof}
Existence of $\phi_t$ for all $t>0$ follows from boundedness. If $y_0=0$, then \eqref{dot_xy} can immediately be integrated  yielding $\phi_t(x_0,0) =(r_0+(x_0-r_0)e^{-at},\,0)$. If 
$r_0\leq 1$, then the disease free equilibrium 
$\bp^*_{dfe}:=(r_0,0)$ is the only equilibrium point (EP) in 
$\P$ and the statement follows by absence of periodic solutions and the Poincaré-Bendixson Theorem. If $r_0>1$, then there also exists the endemic EP 
$\bp^*_{end}:=(1,a(r_0-1))\in\P$ and, by the same argument, the omega limit set of $\{\phi_t(\bp_0)\}$ must consist of one of the two EPs. If $y_0>0$ it must be $\bp^*_{end}$, either by arguing that 
$\bp^*_{dfe}$ is a saddle point with attractive line $\{y=0\}$ (calculate the Jacobian), or by using that 
$$
L(x,y):=y\exp\left[\frac{y+(x-1)^2/2}{a(1-r_0)}\right]
$$
provides a Lyapunov function satisfying $L(x,0)=0$, 
$L(x_0,y_0)>0$ and 
$$
\ga\inv\dot{L}=\frac{(a+y)(x-1)^2}{r_0-1}L\geq 0.
$$
\end{proof}
To study the asymptotic behavior at these equilibria one has to compute eigenvalues and slopes of eigenvectors of the Jacobian. For example, as already noted by Hethcote in \autocite{Hethcote1976, Hethcote1989, Hethcote2000}, see also chapter 3.4-3.5 in the text book by M. Martcheva \autocite{Martcheva}, there is a sub-range $a<1$ and $1<r_-<r_0<r_+$, where the endemic equilibrium becomes spiral and hence this model shows endemic oscillations. A complete detailed analysis of possible asymptotic scenarios has also been given in \autocite{Nill_Omicron}.

\section{The 10-parameter SI(R)S model\label{Sec_SIRS-model}}
\begin{figure}[ht!]
\centering
\includegraphics[width=0.5\textwidth]
	{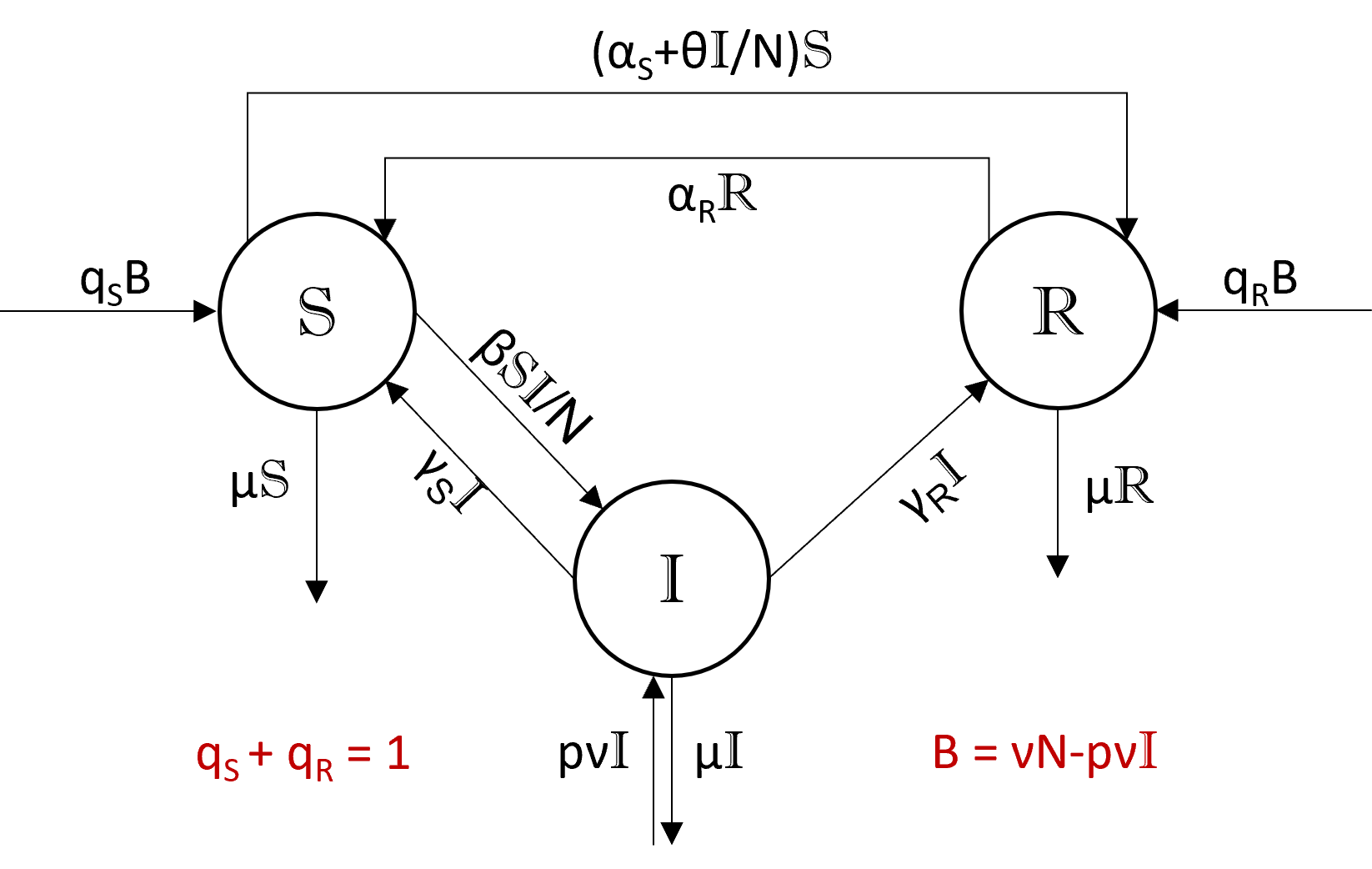}
	\caption{Flow diagram of the SI(R)S model. 
	$B=\nu N-p\nu \II$ denotes the 
	number of not infected newborns per unit of time. 
	}
	\label{Fig_SIRS-Flow}
\end{figure}

This section introduces a homogeneous 10-parameter SI(R)S-model (i.e. mixed SIRS/SIS model) with standard incidence and flow diagram as depicted in Fig. \ref{Fig_SIRS-Flow}. The model describes the infection dynamics of three compartments with populations 
$\bP=(\SS,\RR,  \II)\in\RRN^3$ and total population 
$N=\SS+\II+\RR>0$. Members of $\II$ are infectious, members of 
$\SS$ are susceptible (not immune) and members of 
$\RR$ are immune due to recovery or vaccination.
To model widely experienced social behavior, Fig. \ref{Fig_SIRS-Flow} introduces the parameter 
$\theta$ to the classic setting. It describes the {\em willingness to get vaccinated} by assuming a vaccination rate $\theta I$ proportional to the prevalence $I:=\II/N$. 
As we will see, such an extended model can always be transformed to the standard case $\theta=0$ (Corollary \ref{Cor_theta}).
%

Parameters in this model are

\smallskip
\begin{tabular}{lcp{0.75\textwidth}}
$\al_S$ &:& 
		Constant vaccination rate.
\\
$\theta$ &:& 
		Willingness to get vaccinated given the actual prevalence 
		$I=\II/N$. 
\\
$\al_R$ &:& 
		Immunity waning rate.
\\
$\be$ &:& 
		Effective contact rate of a 
		susceptible from $\SS$.
\\
$\ga_S,\ga_R, \ga$ &:& 
		Recovery rates for $\II\rto \SS$ and $\II\rto \RR$, 
		respectively. $\ga:=\ga_S+\ga_R$.
\\
$\mu$ &:& 
		Mortality rate, assumed to be compartment independent.
\\
$\nu$ &:& 
		Rate of newborns, assumed to be compartment independent.
\\&&	Newborns from 
		$\SS$ and $\RR$ are not supposed to be infected. 
\\
$p$  &:& 
		Probability of a newborn from $\II$ to be infected.
\\
$B$	&:&
		Sum of not infected newborns, 	$B=\nu N -p\nu\II$. 
\\
$q_S, q_R$  &:& 
		Split ratio of not infected newborns landing in
		$\SS$ and $\RR$, 
		$q_S+q_R=1$. So, $q_R$ is the vaccinated portion of not 	
		infected newborns.
\end{tabular}

\medskip\noindent
Epidemiologically all parameters are assumed nonnegative. Also,
$p\leq 1$, $q_S+q_R=1$, $\be>0$ and $\ga>0$. So in total we have 10 parameters, four of which, $(\nu,\mu,p,q_R)$, are purely demographic. Subcases of this model with constant population and $\theta=0$ have been analyzed e.g. in \autocite{KorobWake, ORegan_et_al}.
Of course, Hethcote's classic endemic model is also a special subcase, which has been reinvented several times, see e.g.  \autocite{Ullah_et_al, Chauhan_et_al, Batistela_et_al}. 

At this place one should mention that there are various models in the literature treating vaccination and loss of immunity differently. For example, one might introduce a separate compartment $V$ to distinguish vaccinated from recovered individuals 
(see, e.g. \cite{Arino_et_al}). Time dependent vaccination rates have also been studied in \autocite{LedzSch} by applying optimal control methods and in \autocite{Kopfova_et_al} by letting the vaccination activity be functionally dependent on the history of the prevalence  via the Preisach hysteresis operator. 
Following a similar philosophy, in \autocite{dOnofrio_et_al_2007, dOnofrio_et_al_2008} the authors
use an information variable $M$ to 
model how information on current and past states of the disease influences decisions in families whether to vaccinate or not their children. 

Partial and/or waning immunity may also be modeled  by introducing 
a diminished transmission rate directly from $R$ to $I$ (or $V$ to $I$). Such models are well known to lead to a so-called {\em backward bifurcation}, see e.g. 
\autocite{Had_Cast, Had_Dries, KribsVel, Arino_et_al, AvramAdenane2022, AvramAdenane_et_al}. In fact, the methods of this paper will generalize to such a setting, see \cite{Nill_Symm1}.

\bsn
I am now going to show that the full 10-parameter model in Fig. \ref{Fig_SIRS-Flow} in fact boils down to the extended Hethcote model as defined in Definition \ref{Def_Heth}.

\subsection{Redundancy of birth and death rates}
In a first step, we follow the strategy of \autocite{Nill_Redundancy}, showing that in the dynamics of fractional variables 
$\bp:=N\inv\bP\equiv(S,R,I)$ the four demographic parameters 
$(\mu, \nu, p, q_S=1-q_R)$ become redundant. 
We have $\dot{N}=(\nu-\mu)N$ and  
\begin{equation}
\dot{\bp}=\bL\bp+\bM\bp + \bLa(\bp)+(\mu-\nu)\bp\,,
\end{equation}
\begin{align}
\bL	&:=
\begin{pmatrix}
q_S\nu-\mu	& q_S\nu	& q_S(1-p)\nu
\\
q_R\nu	& q_R\nu-\mu	& q_R(1-p)\nu
\\
0	&0	& p\nu-\mu
\end{pmatrix}
\label{L}
\\
\bM	&:=
\begin{pmatrix}
-\al_S & \al_R & \ga_S
\\
\al_S & -\al_R & \ga_R
\\
0 & 0 & -\ga
\end{pmatrix},
\qquad
\bLa(\bp) :=
\begin{pmatrix}
-(\be+\theta)SI
\\
\theta SI
\\
\be SI
\end{pmatrix}
\end{align}
Now, demographic parameters become redundant by putting 
 $\tilde{M}_{ij}:=M_{ij}+L_{ij}+(\mu-\nu)\delta_{ij}$, i.e.
%
\begin{equation}
\begin{array}{rclrcl}
\tal_S &:=&\al_S+q_R\nu\,,\qquad\qquad\qquad &
\tga_S &:=&\ga_S+q_S(1-p_I)\nu\,,
\\
\tal_R &:=&\al_R+q_S\nu\,, &
\tga_R &:=&\ga_R+q_R(1-p_I)\nu\,.
\end{array}
\label{tilde_parameters} 
\end{equation} 
In this way, the number of effective parameters reduces from ten to six. So, from now on, we put without loss $\mu=\nu=0$ and omit the tilde above parameters. Again, it is convenient to introduce dimensionless parameters. Put $\ga:=\ga_R+\ga_S$ and
\begin{align*}
\al_1 &:=\al_S/\ga, &	\ga_1&:=\ga_S/\ga, &	\be_1:=\be/\ga,
\\
\al_2 &:=\al_R/\ga, &	\ga_2&:=\ga_R/\ga, &	\th_1:=\th/\ga.
\end{align*}
Note that $\be_1\equiv r_0$. The new notation indicates a possible generalization to models where also $R$ is susceptible \cite{Nill_Symm1}. With this notation the dynamics takes the form
\begin{equation}
\ga\inv\dot{\bp}=\bF_\bfa(\bp):=\bM\bp+\bLa(\bp)
\label{dot_bp}
\end{equation}
\begin{align}
\bM	:=
\begin{pmatrix}
-\al_1 & \al_2 & \ga_1
\\
\al_1 & -\al_2 & \ga_2
\\
0 & 0 & -1
\end{pmatrix},
\qquad
\bLa(\bp) :=SI
\begin{pmatrix}
-(\be_1+\theta_1)
\\
\theta_1
\\
\be_1
\end{pmatrix}.
\label{M+Lambda}
\end{align}
We start with an extended parameter space by putting $a:=\al_1+\al_2$, $\be_+:=\be_1+\theta_1$ and requiring 
$\bfa:=(\al_1,\al_2,\ga_1,\ga_2,\be_1,\be_+)\in\A$. Here we put

\begin{align}
\A	&:=\{\bfa\in\RR^4\times\RR_+^2\mid a>0\,\land\,
\ga_1+\ga_2=1\},
\label{A}\\
\Abio	&:=\A\cap\{\al_i\geq 0\,\land\,\ga_i\geq 0\,\land\,
\be_+\geq \be_1\},
\label{Abio}\\
{\Abio}_{,0}	&:=\Abio\cap\{\theta_1=0\},
\label{Abio0}
\end{align}
while $\ga>0$ is understood throughout. So, $\Abio$ denotes the epidemiologically admissible subset of parameters. Also, we start with defining  the system \eqref{dot_bp} on the extended phase space $\bp\in\P$ given by the half plane
\begin{equation}
\P:=\{\bp\equiv(S,R,I)\in\RR^2\times\RRN\mid S+R+I=1\}.
\label{P}
\end{equation}
Clearly, $\P$ stays invariant under the dynamics \eqref{dot_bp} for all $\bfa\in\A$.
Epidemiologically the system is considered for initial conditions in the {\em physical triangle} $\Tph\subset\P$
\begin{equation}
\Tph:=\P\cap\RRN^3.
\label{Tph}
\end{equation}
It is straightforward to check that for all $\bfa\in\Abio$  this triangle  stays forward invariant under the dynamics \eqref{dot_bp}, i.e.  if $\bp\in\Tph$ and $p_i=0$ then
$\dot{p_i}\geq 0$.\footnote
{More generally it can be shown that $\Tph$ is forward invariant iff 
$\bfa\in\A\cap\{\theta_1\geq -\al_1-\ga_2-2\sqrt{\al_1\ga_2}\}$.\label{Fn_Tph}}
 
\begin{remark}
The ``quasi-SIR limit'' $\al_1=\al_2=0$ becomes an integrable Hamiltonian model, see Section \ref{Sec_quasi-SIR} and Appendix \ref{App_alpha=0}.
\end{remark}

\subsection{Dynamics of the replacement number\label{Sec_dot-xy}}
From now on we substitute $R=1-S-I$ and drop the variable $R$. Hence 
$\P=\{(S,I)\in\RR\times\RRN\}$ and 
$\Tph=\{(S,I)\in\RRN^2\mid S+I\leq 1\}\subset\P$. In a second step we now proceed similar as in \cite{Nill_Omicron} and define

\begin{equation}
\begin{aligned}
x&:=\be_1 S,	& y&:=\be_+I,
\\
R_0 &:=\be_1\frac{\al_2}{a}\equiv r_0\frac{\al_2}{a},	& 
\rho&:=(\al_2-\ga_1)\frac{\be_1}{\be_+}.
\end{aligned}
\label{f}
\end{equation}
Then $(x,y)\in\P$ and
the equations of motion \eqref{dot_bp} are equivalent to
\begin{equation}
\begin{aligned}
\ga\inv\dot{x}	&=-xy- \rho y + a(R_0-x),
\\
\ga\inv\dot{y} &=(x-1)y.
\end{aligned}
\label{dot_xy_sirs}
\end{equation}
Here, $x$ is again the replacement number and $R_0$ is the well known {\em vaccination reduced reproduction number} \cite{Driessche2017}. 
Note that for $\bfa\in\Abio$ the definitions imply $0\leq R_0\leq r_0$ and $-1\leq\rho\leq a$. 
Also note that choosing the variables $(x,y)$ has further reduced the number of free parameters from six to four, i.e. the dynamics in \Eqref{dot_xy_sirs} is independent of $\be_1$ and $\be_+$. Instead, these parameters now fix  the image of the physical triangle $\Tph$ in $xy$-space.
\begin{equation}
\Tph(\be_1,\be_+)
:=\{(x,y)\in\RRN^2\mid x/\be_1+y/\be_+\leq 1\},\qquad 0<\be_1\leq\be_+.
\label{Tph(beta)}
\end{equation}
So, given $(a,R_0,\rho)$ and the position of $\Tph\subset\P$, one recovers 
$\bfa\in\A$. More precisely, we have
\begin{lemma}\label{Lem_f}
Put $b:=(\al_2-\ga_1)\be_1\equiv\be_+\rho$ and
$\B:=(\RR_+\times\RR^2)\times\RR_+^2$.
The map 
$$
f:\A\ni\bfa\mapsto(a,R_0,b)\times(\be_1,\be_+)\in\B
$$
is bijective with inverse $f\inv$ given by 
\begin{equation}
\al_1=a-aR_0/\be_1,\qquad\al_2=aR_0/\be_1,\qquad
\ga_1=(aR_0-b)/\be_1, \qquad\ga_2=1-\ga_1.
\label{f_inv}
\end{equation}
Moreover,
\begin{equation}
f(\Abio)=\B\cap\{0\leq R_0\leq\be_1\leq\be_+\,\land\,
aR_0-\be_1\leq b\leq aR_0\}.
\label{f_image}
\end{equation}
\end{lemma}
\begin{proof}
\Eqref{f_inv} is straightforward and the conditions in the r.h.s. of \eqref{f_image} are equivalent to, respectively, 
$0\leq\al_2\leq a$, $\theta_1\geq 0$ and $0\leq\ga_1\leq 1$.
\end{proof}
\noindent
Lemma \ref{Lem_f} motivates the following definition.

\begin{definition}\label{Def_admissible}
Let $a>0$, $0\leq R_0$ and $-1\leq \rho\leq a$ be given. Then 
$(\be_1,\be_+)$, respectively triangles $\Tph(\be_1,\be_+)$, are called {\em admissible}, if
$(a,R_0,\be_+\rho)\times(\be_1,\be_+)\in f(\Abio)$.
\end{definition}
\noindent
Since for $\bfa\in\Abio$ physical triangles are forward invariant under the dynamics \eqref{dot_bp}, we conclude
\begin{corollary}\label{Cor_admissible}
Admissible triangles $\Tph(\be_1,\be_+)$ are forward invariant under the dynamics \eqref{dot_xy_sirs}.
\end{corollary}


\begin{remark}
I should remark that Eqs. \eqref{dot_xy_sirs} have been obtain in equivalent form by Korobeinikov and Wake in \cite{KorobWake}, with 
$\rho\geq 0$, but without the upper bound $\rho\leq a$. This is due to the fact that the authors consider possibly unbalanced birth and death rates, $\mu\neq\nu$, but still require the total population to be time independent. With other words, the recovered/immune compartment is forced to obey
$\dot{\RR}=-\dot{\SS}-\dot{\II}$ anyhow. For $\mu\neq\nu$, this leads to a non-constant and $S$- and $I$-dependent mortality rate in the $R$-compartment. Nevertheless, this system transforms to the present setting, with $\bfa\in\A$, but possibly with $\al_1$ and 
$\ga_2$ negative, so $\bfa\not\in\Abio$, see Appendix \ref{App_KorW} for the details. The fact that global stability results as in \cite{KorobWake} may also hold outside of $\Abio$ will be covered by Theorem \ref{Thm_SIRS} below.
\end{remark}

\subsection{Equilibrium states\label{Sec_Equil}}
From the dynamics \eqref{dot_xy_sirs} we immediately read off the solutions of $\dot{x}=\dot{y}=0$, yielding a disease free equilibrium $(x^*\dfe,y^*\dfe)$ and an endemic equilibrium 
$(x^*\en,y^*\en)$,
\begin{equation}
(x^*\dfe,y^*\dfe)=(R_0,0),\qquad(x^*\en,y^*\en)=(1,a(R_0-1)/(\rho+1)),
\end{equation}
where the endemic equilibrium requires $R_0>1$ and $\rho>-1$.
In terms of original variables and parameters this gives
\begin{align}
(S^*\dfe,I^*\dfe)&=\left(\frac{R_0}{r_0},\,0\right)&&=
\left(\frac{\al_R}{\al_S+\al_R},\,0\right),
\label{dfe}\\
(S^*\en,I^*\en)&=\left(\frac{1}{r_0},\,
\frac{\al_R r_0}{(\al_R+\ga_R)r_0+\theta}(1-\frac{1}{R_0})\right)
\hspace{-30pt}&&=
\left(\frac{1}{r_0},\,
\frac{(r_0-1)\al_R-\al_S}{(\al_R+\ga_R)r_0+\theta}\right).
\label{ee}
\end{align}
This generalizes well known results in the literature  \cite{Hethcote1974, Hethcote1976, Hethcote1989, Hethcote2000, KorobWake, ORegan_et_al, Ullah_et_al, Chauhan_et_al, Batistela_et_al} to the case of our present 10-parameter model. 

\begin{remark}\label{Rem_rho=-1}
As already noted, for $\bfa\in\Abio$ we have 
$-1\leq\rho\equiv b/\be_+\leq a$, where the boundary case 
$\rho=-1$ is equivalent to $\al_2=\ga_2=\theta_1=0$. In particular, it also requires $R_0=0$ and $\be_1=\be_+$. Epidemiologically, this case is  uninteresting and excluded in what follows. It will be discussed shortly in Appendix \ref{App_SIS+vacc}.
\end{remark}

\begin{remark}
Typically, vaccination diminishes the reproduction number $R_0$ as in \Eqref{f},
where $\partial R_0/\partial\al_1<0$. This allows quite generally to determine lower bounds on vaccination rates to achieve herd immunity by requiring $R_0\leq 1$, see e.g. the text book \cite{Martcheva}. In contrast, here $R_0$ is independent of  the $I$-linear vaccination rate $\theta_1$. In fact, $\theta_1$ just diminishes $I^*\en$, see \eqref{ee}, and increases the recovered/immune fraction accordingly. But it doesn't influence the value of $S^*\en$, nor the disease free equilibrium, nor the endemic threshold $R_0=1\LRA r_0=1+\al_1/\al_2$. In fact, as we will
see in Corollary \ref{Cor_theta} in Section \ref{Sec_SIRS-Sym},
by a scaling transformation 
$(S,I)\mapsto(\la S,I)$, $\la=\be/(\be+\theta)$, the SI(R)S model \eqref{dot_bp} with parameters in 
$\bfa\in\Abio$ and $\theta>0$ maps isomorphically to a system with appropriately transformed parameters $\bfa'\in{\Abio}_{,0}$, i.e. $\theta'=0$, while keeping $R_0$ invariant.
\end{remark}

\subsection{Transformation to the extended Hethcote model}
In the third step, we now apply a rescaling transformation of $x-1$ as first introduced in \cite{Nill_Omicron}. 
This will show that for $\rho>-1$ the system \eqref{dot_xy_sirs} is isomorphic to an extended Hethcote model as defined in Definition
\ref{Def_Heth}. Hence, to prove stability properties for the above  SI(R)S model equilibria, we will just need to quote Hethcote's results in the formulation of Theorem \ref{Thm_Heth}.

\begin{proposition}[Nill 2022 \cite{Nill_Omicron}]\label{Prop_Nill}
Consider the system \eqref{dot_xy_sirs} on phase space $\P$ and for parameters $\ga>0$, $\rho>-1$, $a\geq 0$ and $R_0\in\RR$. Define rescaled variables and parameters by
\begin{equation}
\bar{x}-1=\frac{x-1}{\rho+1},\quad \bar{y}=\frac{y}{\rho+1},
\quad\bar{a}=\frac{a}{\rho+1},\quad
\bar{R}_0-1=\frac{R_0-1}{\rho+1},\quad\bar{\ga}=(\rho+1)\ga.
\label{bar}
\end{equation}
Then $(\bar{x},\bar{y},\bar{\ga},\bar{a},\bar{R}_0)\in\P\times\RR_+^2\times\RR$ and  the system \eqref{dot_xy_sirs} is isomorphic to the extended Hethcote model \eqref{dot_xy}.
\begin{equation}
\bar{\ga}\inv\dot{\bar{x}}	=-\bar{x}\bar{y} + 
\bar{a}(\bar{R}_0-\bar{x}),
\qquad
\bar{\ga}\inv\dot{\bar{y}} =(\bar{x}-1)\bar{y}.
\label{dot_txy_sirs}
\end{equation}
\end{proposition}
\begin{proof}
By straightforward calculation.
\end{proof}
Since $\sign(\bar{R}_0-1)=\sign(R_0-1)$ and 
$\sign(\bar{x}-1)=\sign(x-1)$, the results of Theorem \ref{Thm_Heth} now immediately translate to our original model. In doing so, due to the global boundedness property in Lemma \ref{Lem_Heth}, we no longer have to restrict ourselves to parameter constraints 
$\bfa\in\Abio$ to guarantee forward invariance of physical triangles. The following more general definition of 
$\AH$ will do.

\begin{theorem}\label{Thm_SIRS}
Consider the SI(R)S model \eqref{dot_bp} on phase space $\P$ \eqref{P} with parameters 
\begin{equation}
\bfa\in\AH:=\A\cap\{\rho>-1\}\equiv
\A\cap\{\theta_1/\be_1>-\al_2-\ga_2\}.
\label{A+}
\end{equation}
\begin{itemize}
\item[i)]
For any initial conditon $\bp_0=(S_0,I_0)\in\P$ the forward orbit 
$\phi_t(\bp_0)$ exists for all $t>0$. If $R_0\leq 1$ or $I_0=0$, then 
$
\lim_{t\rto\infty}\phi_t(\bp_0)=(S^*\dfe,I^*\dfe).
$
Otherwise 
$\lim_{t\rto\infty}\phi_t(\bp_0)=(S^*\en,I^*\en).
$
\item[ii)]
If $\bfa\in\AH\cap\{\al_i\geq 0\}$ then $(S^*\dfe,I^*\dfe)\in\Tph$.
\item[iii)]
If $\bfa\in\AH\cap\Abio$
and $R_0>1$, then $(S^*\en,I^*\en)\in\Tph$.
\end{itemize}
\end{theorem}
\begin{proof}
Under the transformations \eqref{f} the equivalence of systems \eqref{dot_bp} and \eqref{dot_xy_sirs} holds for all 
$\bfa\in\A$ and the equivalence of \eqref{dot_xy_sirs} and \eqref{dot_txy_sirs} holds for all $\rho>-1$ by Proposition \ref{Prop_Nill}. Part i) follows from Theorem \ref{Thm_Heth}, part ii) is obvious from $0\leq\al_i/a\leq 1$ and part iii) follows from forward invariance of $\Tph$ for 
$\bfa\in\Abio$.
\end{proof}

\subsection{The quasi SIR limit\label{Sec_quasi-SIR}}
In the limit $\al_1=\al_2=0$, $\ga_i\geq 0$ and $\theta_1\geq 0$  the system \eqref{dot_bp} becomes a combined SIR/SIS model with absence of immunity waning and just an $I$-linear vaccination rate. In this case, we get $a=aR_0=0$ and $\rho=-\ga_1\be_1/\be_+\in[-1,0]$. For $\rho=-1$ this is the pure SIS model, $\ga_2=\theta_1=0$, and for $-1<\rho\leq 0$ the transformation  \eqref{bar} reduces to the classic SIR model, \Eqref{dot_txy_sirs} with $\bar{a}=\bar{a}\bar{R}_0=0$. Hence, this model also shows {\em herd immunity} and standard results for the classic SIR model \cite{KerMcKen, Hethcote1970, Hethcote1976, Hethcote1989, HLM, Prodanov} now apply, see Appendix \ref{App_alpha=0} for more details. Below let  
$W_+:[-1/e,\infty)\to[-1,\infty)$ denote the upper branch of the 
so-called Lambert-W function \cite{Wolfram, Lambert-W}, i.e. the inverse of 
$[-1,\infty)\ni z\mapsto ze^z\in[-1/e,\infty)$. 

\begin{theorem}\label{Thm_a=0}
Consider the SI(R)S model \eqref{dot_bp} on the physical triangle 
$\Tph$ \eqref{Tph} for $\al_i=0$, $0\leq\ga_i\leq 1$ and 
$\theta_1\geq 0$, excluding the case $(\ga_1,\theta_1)=(1,0)$ (pure SIS model). Assume  
$\be_1\equiv r_0>1$ and initial conditions $(S_0,I_0)\in\Tph$, $I_0>0$
and $S_0>\ga_1/(\be_1+\theta_1)$.
\begin{itemize}
\item[i)]
The limits $\lim_{t\to\pm\infty}(S(t),I(t))=(S_{\pm\infty},0)$ exist and satisfy 
\begin{equation}
\ga_1/(\be_1+\theta_1)<S_\infty<r_0\inv<S_{-\infty}<\infty.
\label{S_infty}
\end{equation}
\item[ii)]
Put $\rho=-r_0\ga_1/(\be_1+\theta_1)$ and 
$\bar{x}_{-\infty}=(r_0S_{-\infty}+\rho)/(\rho+1)$.
Then the following generalized final size formula holds
\begin{equation}
r_0S_\infty=
-(\rho+1)W_+\left(-\bar{x}_{-\infty}e^{\bar{x}_{-\infty}}\right)-\rho.
\label{final_size}
\end{equation}
\item[iii)] 
Assume $S_{-\infty}\leq 1$. As $t\to\+\infty$ a fraction $\theta_1/(\theta_1+r_0\ga_2)\,\Del R$ of the total increase 
$\Del R=S_{-\infty}-S_{+\infty}$ in the $R$-compartment is vaccinated.
\end{itemize}
\end{theorem}
\noindent

\begin{remark}
Note that $\Tph$ is not backward invariant, i.e. depending on 
$(S_0,I_0)$ one may possibly get $S_{-\infty}>1$. Also note that the case $\rho=0\LRA\ga_1=0$ reduces to the classic SIR model for variables $(S,I)$, equipped with an $I$-linear vaccination rate $\theta=\ga\theta_1$. In this case, the final size \eqref{final_size} is independent of $\theta_1$. In fact, under the usual assumption 
$S_{-\infty}=1$, it reduces to the standard final size formula in the classic SIR model, see e.g.
\cite{KerMcKen, Hethcote1970, Hethcote1976, Hethcote2000, MaEarn, Miller, Weiss}. Theorem \ref{Thm_a=0} is proven in Appendix \ref{App_alpha=0}, where also the cases 
$S_0\leq\ga_1/(\be_1+\theta_1)$ and $r_0\leq 1$ are discussed.
\end{remark}

In summary, in this section we have seen that, for parameters 
$\bfa\in\AH$ \eqref{A+}, the SI(R)S model \eqref{dot_bp} is isomorphic to the extended Hethcote model \eqref{dot_txy_sirs}, and that in the limit $\al_i=0$ and $\rho+1\in(0,1]$ we get a quasi-SIR model. The equivalences of these models have been obtained by applying  three scaling transformations
\begin{equation}
S\rto x=\be_1 S,\qquad
x-1\rto\bar{x}-1=(x-1)/(\rho+1), \qquad I\rto \bar{y}=\be_+I/(\rho+1).
\label{new_variables}
\end{equation}
The next section studies these transformations and the relation between $\AH$ and $\Abio$ more systematically under group theoretical aspects.

\section{Symmetries and parameter reduction\label{Sec_symmetries}}
\subsection{Basic concepts\label{Sec_basics}}
For simplicity, unless stated explicitly, all maps in this section are supposed to be $\C^{\infty}$.
A {\em model class} on some phase space $\P$ is a family of dynamical systems $\dot{\bp}=\bF(\bp,\bfa)$, where 
$\bF_\bfa:=\bF(\cdot,\bfa)\in\Omega^1(\P)$ are  vector fields on $\P$ parametrized by a set of external parameters $\bfa\in\A$.
Typically  $\A\subset\RR^m$ and for $\bfa=(a_1,\cdots,a_m)\in\A$ we have $\bF=\bF_0+\sum_{\nu=1}^m a_\nu\bF_\nu$, where
$\bF_\nu\in\Omega^1(\P)$, $1\leq\nu\leq m$, are linearly independent as vector fields on $\P$. 
Given a model class $\bF$, it is helpful to consider $\bfa\in\A$ also as dynamical variables obeying $\dot{\bfa}=\bzero$. Putting 
$\M:=\P\times\A$, the associated vector field 
$\bF_\M\in\Omega^1(\M)$ is given by
$\bF_\M(\bp,\bfa):=(\bF(\bp,\bfa),\bzero)\in
T_{\bp}P\oplus\bzero_{T_{\bfa}\A}\subset T_{(\bp,\bfa)}\M$.

Two model classes $(\P,\A,\bF)$ and $(\tP,\tA,\tbF)$ are said to be isomorphic, if there exists a diffeomorphism 
$\Phi:\M=\P\times\A\rto\tilde{\M}=\tP\times\tA$ projecting to a diffeomorphism $\varphi:\A\rto\tA$ (i.e. 
$\pi_{\tA}\circ\Phi=\varphi\circ\pi_\A$), such that 
$\Phi_*\bF_\M=\tbF_{\tilde{\M}}$.\footnote{Here, $\pi_\A:\M\rto\A$ denotes the canonical projection and, for diffeomorphisms $\psi:\M\rto\N$, 
$\psi_*:\Omega^1(\M)\rto\Omega^1(\N)$ denotes the ``push forward'' on vector fields, 
$\psi_*\bF:=\bD\psi\circ\bF\circ\psi\inv$.}

\begin{definition}\label{Def_symmetry}
A {\em parameter symmetry group} 
$\G$ of a model class $\bF$ is a left $\G$-action on $\M=\P\times\A$ by diffeomorphisms $\bpsi_g:\M\to\M$, $g\in\G$, 
projecting to a $\G$-action on $\A$, denoted by 
$\G\times\A\ni(g,\bfa)\mapsto g\re\bfa\in\A$, such that 
$(\bpsi_g)_*\bF_\M =\bF_\M$ for all $g\in\G$.
\end{definition}

\ssn
To understand this definition, let 
$\bpsi_g(\bp,\bfa)=(\psi_{g}(\bp,\bfa),g\re\bfa)$. 
Then 
$
\bD_\M\bpsi_g(\bp,\bfa)\circ\bF_\M(\bp,\bfa)=
(\bD_\P\psi_g(\bp,\bfa)\circ\bF_\bfa(\bp),\bzero)
$
and $\G$ acts as a parameter symmetry group, iff for all $g\in\G$ and $\bfa\in\A$ 
\begin{equation}
\bD_\P\psi_g(\bp,\bfa)\circ\bF_\bfa(\bp)=
\bF_{g\re\bfa}(\psi_g(\bp,\bfa)).
\label{psi_g}
\end{equation}
In other words, the equations of motion stay invariant under the transformation $\psi_{g}(\cdot,\bfa):\P\rto\P$, if we transform parameters $\bfa\in\A$ accordingly. Also note that in most common examples the $\G$-action on $\M$ factorizes,
i.e. $\psi_g$ is independent of $\bfa$ and $\bpsi_g(\bp,\bfa)=(\psi_g(\bp),g\re\bfa)$.

Given a parameter symmetry $\G$, the family of parametrized dynamical systems 
$\dot{\bp}=\bF_\bfa(\bp)$ falls into isomorphy classes labeled by the orbits $[\bfa]\equiv\G\re\bfa\in\A/\G$. Under some technical assumptions, this allows to construct equivalent transformed systems with reduced parameter space $\A/\G$. Equivalently, one may  ``turn parameter space north'' by choosing a suitable section 
$\sigma:\A/\G\rto\A$ and solving the system for 
$\bfa\in\sigma(\A/\G)$.
A simple example would be $\P=\RR^n$, $\dot{\bp}=\bA\bp$, 
$\bA\in\A$ the space of symmetric $n\times n$-matrices, and 
$\G=SO(n)$. In this case parameter reduction means diagonalizing 
$\bA$. Another example would be the dynamics of mutually interacting classical particles  in a constant external (say magnetic) field 
$\bB=(B_1,B_2,B_3)$.
In this case $\A=\RR^3\setminus\{\bzero\}$ and $\G=SO(3)$ and ``turning parameter space north'' means putting without loss
$g\re\bB=(0,0,|\bB|)$. As a third example, also quasimonomial transformations to canonical forms for generalized Lotka-Volterra (GLV) systems can be understood in this way, see \autocite{Brenig, Hernandez_Fairen}.

\subsection{The SI(R)S symmetry\label{Sec_SIRS-Sym}}
We now apply this formalism to the 6-parameter SI(R)S model \eqref{dot_bp} with phase space $\P\cong\RR\times\RRN$ and parameter space 
$\hA:=\{(\ga,\bfa)\in\RR_+\times\A\}$, where $\A$ is given in \Eqref{A}. Following the three scaling transformations in \eqref{new_variables}, denote $\G$ the multiplicative abelian group
$$
\G=\GS\times\GI\times\GX=\RR_+^3.
$$
Elements of $\G$ are denoted $g=(\lambda,\eta,\xi)\in\RR_+^3$. By convenient notation, we identify 
$\lambda\equiv(\lambda,1,1)\in\GS\subset\G$ etc. Put $\M=\P\times\hA$ and $\N=\P\times\hB$, where $\hB=\RR_+\times\B$ and where 
$\B=(\RR_+\times\RR^2)\times\RR_+^2$ has been defined in Lemma \ref{Lem_f}. To define the action 
$\bpsi_g:\M\rto\M$ in the sense of Definition \ref{Def_symmetry} we first transform to an isomorphic model class  by applying the diffeomorphism
\begin{equation}
\Phi:\M\ni(S,I,\ga,\bfa)\mapsto
\left(u:=\be_1S-1,\,I,\,\ga,\,f(\bfa)\right)\in\N,
\label{Phi}
\end{equation}
where $f:\A\rto\B$ has been defined in Lemma \ref{Lem_f}. The $\Phi$-transformed dynamical system with phase space 
$(u,I)\in\P$ and  parameters 
$(\ga,a,R_0,b,\be_1,\be_+)\in\hB$ is then obtained by replacing $y=\be_+ I$ and $x=u+1$ in \eqref{dot_xy_sirs}.
\begin{equation}
\begin{aligned}
\ga\inv\dot{u}	&=
-\be_+uI -au- (b+\be_+) I + a(R_0-1)
\\
\ga\inv\dot{I} &=uI.
\end{aligned}
\label{dot_xI_sirs}
\end{equation}
Hence, using $(a,R_0-1,b+\be_+,\be_1,\be_+)$ as adapted coordinates in $\B$, the scaling symmetry $\G$ operates linearly and factorizing on 
$\N$. 
\begin{theorem}
For $g=(\la,\eta,\xi)\in\G$ and 
$\bfb:=(a,R_0-1, b+\be_+,\be_1,\be_+)\in\B$ let the $\G$-action 
$\reB:\G\times\B\rto\B$ be given by 
\begin{equation}
g\reB \bfb:=
\left(\xi a,\,\xi(R_0-1),\,
\eta\inv\xi^2 (b + \be_+),\,\la\inv\xi\be_1,\,\eta\inv\xi\be_+\right)
\label{reB}
\end{equation}
and put $\bL_g:\N\rto\N$,
\begin{equation}
\bL_g(u,I,\ga,\bfb):=
\left(\xi u,\,\eta I,\,\xi\inv\ga,\,g\reB\bfb\right).
\label{Lg}
\end{equation}

\begin{itemize}
\item[i)]
Then $\bL_g$ provides a parameter symmetry of the model class \eqref{dot_xI_sirs}.
\item[ii)]
Put $\bpsi_g:=\Phi\inv\circ\bL_g\circ\Phi:\M\rto\M$. Then $\bpsi_g$ provides a parameter symmetry of the SI(R)S model \eqref{dot_bp} satisfying
\begin{equation}
\bpsi_g(S,I,\ga,\bfa)=
\left(\la S+\la\be_1\inv(\xi\inv-1),\,\eta I,\, \xi\inv\ga,\, g\reA\bfa\right).
\end{equation}
Here the $\G$-action $\reA:\G\times\A\rto\A$ is given by  $g\reA:=f\inv\circ g\reB\circ f$.
\item[iii)]
For 
$\bfa=(\al_1,\al_2,\ga_1,\ga_2,\be_1,\be_+)^T\in\A$ 
and $\bC\bfa:=(a,-a,-a-\be_+,a+\be_+-\be_1,0,0)^T\in\RR^6$, $a=\al_1+\al_2$,  we have 
\begin{align}
g\reA\bfa	&=
\bD_S(\la)\bD_I(\eta\inv)\left(\xi\one+\frac{\xi-1}{\be_1}\bC\right)\bfa,
\label{reA}
\\[5pt]
\bD_S(\la)&:=
\begin{pmatrix}
1&1-\la&0&0&0&0
\\
0&\la&0&0&0&0
\\
0&0&\la&0&0&0
\\
0&0&1-\la&1&0&0
\\
0&0&0&0&\la\inv&0
\\
0&0&0&0&1-\la\inv&1
\end{pmatrix},
\label{DS}
\\[5pt]
\bD_I(\eta)&:=
\begin{pmatrix}
1&0&0&0&0&0
\\
0&1&0&0&0&0
\\
0&1-\eta&\eta&0&0&0
\\
0&\eta-1&1-\eta&1&0&0
\\
0&0&0&0&1&0
\\
0&0&0&0&\eta-1&\eta
\end{pmatrix}.
\label{DI}
\end{align}
\end{itemize}
\end{theorem}

\begin{proof}
Parts i) and ii) are obvious. 
(Note that invariance under the action of $\la\in\GS$ in \eqref{reB} follows trivially, since the dynamics in 
\eqref{dot_xI_sirs} is independent of $\be_1$.)
To prove part iii), since we already know that by construction $\reA$ provides a $\G$-action on $\A$, it suffices to prove the formulas  \eqref{reA}-\eqref{DI} separately for $g=\la$, $g=\eta$ and 
$g=\xi$. This is a straight forward calculation, which is left to the reader.
\end{proof}

As a particular consequence we now get, that SI(R)S models with an $I$-linear vaccination $\theta$ always map isomorphically to models with a constant vaccination, 
$\theta'=0$.  

\begin{corollary}\label{Cor_theta}
For $\bfa=(\al_1,\al_2,\ga_1,\ga_2,\be_1,\be_+)\in\Abio$ put 
$\la:=\be_1/\be_+\leq 1$ and 
$\bfa':=\bD_S(\la)\bfa$. Then 
\begin{equation}
\bfa'=(a-\la\al_2,\la\al_2,\la\ga_1,1-\la\ga_1,\be_+,0)\in{\Abio}_{,0}
\label{theta'=0}
\end{equation}
and the scaling transformation 
$(S,I)\mapsto(\la S,I)$ maps the SI(R)S model with parameters $\bfa$ isomorphically to the model with parameters $\bfa'$ while keeping  $R_0'=R_0$.
\qed
\end{corollary}

\begin{remark}
While in general $\psi_g(\cdot,\ga,\bfa)$ will not preserve physical triangles and $g\reA$ will not preserve $\Abio$, $\la\leq 1$ in Corollary \ref{Cor_theta} assures that both statements hold for $g=\la$.
\end{remark}

\noindent
The fact that the $\G$-action does not preserve the physical triangle also implies that for $\bfa\in\AH\setminus\Abio$ disease free or endemic equilibria may well lie outside $\Tph$.

\begin{corollary}\label{Cor_equil>1}
Let $(S^*(\bfa),I^*(\bfa))$ be a disease free or endemic equilibrium,
\eqref{dfe}-\eqref{ee}. Then 
$$
(S^*(g\reA\bfa),I^*(g\reA\bfa))=
(\la S^*(\bfa)+\la\be\inv(\xi\inv-1),\eta I^*(\bfa)).
$$ 
In particular, for all $\bfa\in\AH$ there exists 
$\tilde{\bfa}=g\reA\bfa\in\AH$ \footnote{Here we anticipate $\G\reA\AH=\AH$, see Lemma \ref{Lem_triv}.} such that 
$S^*(\tilde{\bfa})+I^*(\tilde{\bfa})>1$.
\qed
\end{corollary}

\subsection{Parameter reduction\label{Sec_reduction}}
This subsection gives a group theoretical approach to the parameter reduction in Proposition \ref{Prop_Nill} and Theorem \ref{Thm_SIRS}. First we look at the parameter subspace $\AH\subset\A$ introduced in Theorem \ref{Thm_SIRS}.
\begin{lemma}\label{Lem_triv}
Let $\AH\subset\A$ be given by \Eqref{A+}. Then $\G\reA\AH=\AH$ and $\AH\cong\AH/\G\times\G$ as trivial principal $\G$-bundles. A choice of trivialization is given by 
\begin{equation}
\Ga_\A:\AH\ni\bfa\mapsto(\bar{a},\bar{R}_0)\times
(\frac{1}{\bar{\be}_1},\frac{1}{\bar{\be}_+}, \rho+1)
\in(\RR_+\times\RR)\times\RR_+^3\cong\AH/\G\times\G,
\end{equation}
where $\bar{\be}_1=\be_1/(\rho+1)$, 
$\bar{\be}_+=\be_+/(\rho+1)$ and where $(\bar{a},\bar{R}_0)$ have been defined in \eqref{bar}.
\end{lemma}
\begin{proof}
We equivalently prove the statements with $(\AH,\reA,\Ga_\A)$ replaced by 
$(\B^\star,\reB,\Ga_\B)$, where 
$$
\B^\star:=f(\AH)=\{(a,R_0,b,\be_1,\be_+)\in\B\mid b+\be_+>0\}
$$
and where $\Ga_\B:=\Ga_\A\circ f\inv$. 
Using $(\rho+1)\be_+=b+\be_+$, Lemma \ref{Lem_f} and \Eqref{bar}, one immediately checks that
$\Ga_\B:\B^\star\rto(\RR_+\times\RR)\times\RR_+^3$ is a diffeomorphism  with $\Ga_\B\inv$ given by
\begin{equation}
(a,R_0-1, b+\be_+,\be_1,\be_+)= (\rho+1)
\left(\bar{a},\bar{R}_0-1,(\rho+1)\bar{\be}_+,\bar{\be}_1,\bar{\be}_+\right).
\label{Gamma_B_inv}
\end{equation}
Moreover, by \eqref{reB}, 
$\Ga_\B\circ g\reB=(\id_{\RR_+\times\RR}\times\ell_g)\circ\Ga_\B$, where $\ell_g:\G\rto\G$ denotes left multiplication by $g$. Hence 
$\B^\star\cong\B^\star/\G\times\G$ as principal $\G$-bundles.
\end{proof}

In the obvious way, this structure also lifts to 
$\hat{\A}^\star:=\RR_+\times\AH$ with $\G$-action $g\rehA(\ga,\bfa):=(\xi\inv\ga,g\reA\bfa)$ and trivialization 
$$\Ga_{\hat{\A}}:\hat{\A}^\star\ni(\ga,\bfa)\mapsto(\bar{\ga},\Ga_\A(\bfa))
\in \hat{\A}^\star/\G\times\G,
$$
where $\bar{\ga}:=(\rho+1)\ga$, see \eqref{bar}.

Given such a setting, parameter reduction as in \eqref{dot_txy_sirs} is obtained in general by passing from a model class 
$(\P,\A,\bF)$ to an isomorphic model class 
$(\P,\A/\G,\bar{\bF})$ as follows. Any  trivialization 
$\A\cong\A/\G\times\G$ 
is of the form $\Ga_{\A}(\bfa)=([\bfa],h(\bfa))$,
where $h:\A\rto\G$ satisfies $h(g\reA\bfa)=gh(\bfa)$. Putting 
again $\M=\P\times\A$ and denoting 
$\bar{\M}:=\P\times(\A/\G\times\G)$ there is a naturally induced diffeomorphism 
$\bar{\Phi}:\M\rto\bar{\M}$,
$$
\bar{\Phi}(\bp,\bfa):=
\left(\bpsi_{h(\bfa)\inv}(\bp,\,\bfa),\,[\bfa],\, h(\bfa)\right).
$$
The $\bar{\Phi}$-transported $\G$-action, 
$\bar{\bpsi}_g:=
\bar{\Phi}\circ\bpsi_g\circ\bar{\Phi}\inv:\bar{\M}\rto\bar{\M}$, is given by $\bar{\bpsi}_g(\bp,[\bfa],h)=(\bp,[\bfa],gh)$.
Since $(\bpsi_g)_*\bF_\M=\bF_\M$, the transported vector field, 
$\bar{\bF}_{\bar{\M}}:=\bar{\Phi}_*\bF_\M$, is invariant under the transported $\G$-action,
$(\bar{\bpsi}_g)_*\bar{\bF}_{\bar{\M}}=\bar{\bF}_{\bar{\M}}$,
and hence, using \eqref{psi_g} and $\bD_\P\bar{\bpsi}_g=\bone$, 
$\bar{\bF}_{([\bfa],h)}=\bar{\bF}_{([\bfa],gh)}$ for all $g\in\G$. Thus, $\bar{\bF}$ only depends on 
$(\bp,[\bfa])\in\P\times\hat{\A}/\G$.

In our case we have to replace $\A$ by $\hA^\star$ and put
$h(\ga,\bfa)=(\bar{\be}_1\inv,\bar{\be}_+\inv,\rho+1)$. Then 
$$
\bar{\Phi}(\bp,\ga,\bfa)=
(\bar{x},\bar{y})\times(\bar{\ga},\bar{a},\bar{R}_0)\times
({\bar{\be}_1}\inv, {\bar{\be}_+}\inv,\rho+1)
$$
and the $\bar{\Phi}$-transported vector field $\bar{\bF}$ is given by \eqref{dot_txy_sirs} and independent of
$({\bar{\be}_1}, {\bar{\be}_+},\rho+1)$.

\begin{remark}
Note that, similarly as in \eqref{Tph(beta)}, the $\G$-fiber coordinates $({\bar{\be}_1}, {\bar{\be}_+},\rho+1)$ are again precisely determined by the images of physical triangles in 
$(\bar{x}, \bar{y})$-space.
$$
\Tph(\bar{\be}_1,\bar{\be}_+,\rho+1):=
\{\bar{\bp}=(\bar{x}, \bar{y})\in\RR\times\RRN\mid
0\leq\frac{1}{\bar{\be}_1}(\bar{x}-\frac{\rho}{\rho+1})+
\frac{\bar{y}}{\bar{\be}_+}\leq 1\}.
$$
Geometrically these are the triangles with corners 
$\bar{\bp}_\li=(\rho/(\rho+1),0)$, 
$\bar{\bp}_\re=(\rho/(\rho+1)+\bar{\be}_1,0)$ and 
$\bar{\bp}_{\up}=(\rho/(\rho+1),\bar{\be}_+)$.
One may now proceed as in Definition \ref{Def_admissible} and call
$(\bar{\be}_1,\bar{\be}_+,\rho+1)$ (or the triangle
$\Tph(\bar{\be}_1,\bar{\be}_+,\rho+1)$)
{\em admissible} with respect to 
$(\bar{a},\bar{R}_0)$, if 
$\Ga_\A\inv(\bar{a},\bar{R}_0,\bar{\be}_1,\bar{\be}_+,\rho+1)\in\Abio$.
As in Corollary \ref{Cor_admissible}, this implies, that for given 
$(\bar{a},\bar{R}_0)$ admissible triangles are always forward invariant w.r.t. the dynamics \eqref{dot_txy_sirs}. Using Lemma \ref{Lem_f}, it is straight forward to derive conditions for admissibility of $(\bar{\be}_1,\bar{\be}_+,\rho+1)$. Since formulas don't look enlightening, this is left as an exercise to the reader.
\end{remark}

\subsection{Fixing the gauge\label{Sec_gaugefix}}
In physics terminology, ``fixing the gauge'' means choosing a representative from an equivalence class. In our case this may be rephrased by ``turning parameter space north'', i.e. choosing a section $\sigma:\AH/\G\to\AH$.
We now show that for $\bfa\in\AH$ a representative of the equivalence class $[a]\in\AH/\G$ can always be chosen in ${\Abio}_{,0}:=\Abio\cap\{\theta_1=0\}$.

\begin{proposition}\label{Prop_gaugefix}
Let $R_0=\be_1\al_2/(\al_1+\al_2)$ be the vaccination reduced reproduction number as in \eqref{f} and use 
$(\bar{a},\bar{R}_0)\in\RR_+\times\RR$ as global coordinates in $\AH/\G$, see Proposition \ref{Prop_Nill} and  \Eqref{dot_txy_sirs}.
Pick $c>\max\{0,\bar{R}_0,\bar{a}\bar{R}_0\}$ arbitrary.
\begin{itemize}
\item[i)]
An element $\bfa\in\AH$ in the equivalence class
$(\bar{a},\bar{R}_0)$ exists uniquely under the conditions
\begin{itemize}
\item[a)] If $\bar{R}_0>0:
\qquad\theta_1=0,\qquad \be_1=c,\qquad \al_2=\ga_1$.
\item[b)]
If $\bar{R}_0\leq 0:\qquad
\theta_1=0,\qquad \be_1=c,\qquad \al_2=0$.
\end{itemize}
\item[ii)]
Under these conditions $\bfa\in{\Abio}_{,0}\setminus \{\rho=-1\}$ and therefore 
$\AH=\G\reA({\Abio}_{,0}\setminus \{\rho=-1\})$.
\end{itemize}
\end{proposition}
\begin{proof}
In both cases put 
$\be_+=\be_1=c\Longleftrightarrow\theta_1=0$. In case a) 
the condition $\al_2=\ga_1$ implies $\rho=0$ by \Eqref{f} and hence $a=\bar{a}>0$ and $R_0=\bar{R}_0>0$ by
\Eqref{bar}. Next, again by \Eqref{f}, $0<\ga_1=\al_2=aR_0/c\leq\min\{1,a\}$, 
$\ga_2=1-\ga_1$ and $\al_1=a-\al_2$. This proves that $\bfa$ exists uniquely and 
$\bfa\in{\Abio}_{,0}\setminus \{\rho=-1\}$.

In case b) 
$\al_2=0$ implies $R_0=0$ and therefore, by \Eqref{bar},
$\rho+1=(1-\bar{R}_0)\inv\in(0,1]$. Hence,
$\al_1=a=(\rho+1)\bar{a}>0$ and from $\be_1=\be_+=c$ and 
\eqref{f} we conclude
$\ga_1=\al_2-\rho=-\rho$ and $\ga_2=\rho+1$. So, also here $\bfa$ exists uniquely and 
$\bfa\in{\Abio}_{,0}\setminus \{\rho=-1\}$.
\end{proof}

\begin{remark}\label{Rem_gaugefix}
If in Proposition \ref{Prop_gaugefix} 
$c(\bar{a},\bar{R}_0)>\max\{0,\bar{a}\bar{R}_0, \bar{R}_0\}$ is chosen as a smooth function, then the section $\sigma:\AH/\G\to{\Abio}_{,0}$ defined by the above conditions is $C^\infty$ for 
$\bar{R}_0\neq 0$, but only continuous at $\bar{R}_0=0$.
\end{remark}

\begin{remark}\label{Rem_admissible}
Proposition \ref{Prop_gaugefix} may be reformulated by stating that 
$(\bar{\be}_1,\bar{\be}_+,\rho+1)\in\RR_+^3$ are admissible w.r.t. $(\bar{a}, \bar{R}_0)$, if 
$\bar{R}_0>0$, $\rho=0$ and $\bar{\be}_1=\bar{\be}_+=c$, or if
$\bar{R}_0\leq 0$, $\rho+1=(1-\bar{R}_0)\inv$ and 
$\bar{\be}_1=\bar{\be}_+=c(1-\bar{R}_0)$, where 
$c>\max\{0,\bar{R}_0,\bar{a}\bar{R}_0\}$.
\end{remark}

\section{Summary and outlook}
In summary, in this paper I have demonstrated that symmetry concepts in parametrized dynamical systems may help to reduce the number of external parameters by a suitable normalization prescription. If the symmetry group $\G$ is an $n$-dimensional Lie Group and the $\G$-action on parameter space $\A$ admits a trivialization, 
$\A\cong\A/\G\times\G$ as principal $\G$-bundles, then there is a natural diffeomorphism mapping the original system with parameters in $\A$ to an equivalent system with parameter space $\A/\G\times\G$. For the transformed system, invariance under 
$\G$ simply means that the dynamics only depends on $\A/\G$, thus reducing the number of essential parameters by $n$. If, as a principal $\G$-bundle, $\A$ is only locally trivial, this procedure still works by covering $\A$ with suitable charts 
$U\cong U/\G\times\G$.  In an obvious way, this algorithm would also generalize to the case $\A\cong \A/\G\times\V$, where $\G$ acts transitively (but possibly not freely) on the fiber $\V$.

This strategy applies to the fractional dynamics of a general class of epidemic SI(R)S models, with standard incidence and up to ten parameters, including immunity waning, two recovery flows and constant and $I$-linear vaccination rates. Omitting four redundant demographic parameters, 
this model admits $\G\cong\RR_+^3$ as a symmetry group, acting on phase space by rescaling $S,I$ and $x-1$, respectively, $x$ being the replacement number. Thus, identifying the total waiting time $\ga\inv$ in $I$ as a pure time scale, we get a normalized version with essentially just 2 independent parameters, which turns out to be a marginally extended version of Hethcote's classic endemic model  first presented in 1973.

To apply this framework, we had to extend phase space $\P$ by allowing $(S,I)\in\RR\times\RR_+$, while keeping $R=1-S-I$. At the same time, the range of parameters had to be enlarged to $\AH$, including possibly non-physical negative values. As it turned out, apart from an uninteresting boundary case\footnote{see Appendix \ref{App_SIS+vacc}}, $\AH$ coincides with the $\G$-orbit of the epidemiologically admissible parameter subset $\Abio$. Thus, by symmetry arguments, proving endemic bifurcation and stability results in any of these models becomes needless, it's all contained in Hethcote's original work.

Of course, one has to be aware that, for $\bfa\in\AH\setminus\Abio$, equilibrium states may possibly lie outside the physical triangle 
$\Tph=\{(S,I,R)\in\RRN^3\mid S+I+R=1\}$. As shown in Appendix \ref{App_KorW}, although not addressed by the authors, such a scenario may indeed show up in the Korobeinikov/Wake type of SIRS model  \cite{KorobWake}.

As a special consequence, we have also seen that $I$-linear vaccination may always be ``scaled to zero'', i.e. without leaving 
$\Abio$ or $\Tph$ there always exists a $\G$-equivalent system with 
$\theta=0$. In particular, since the threshold for endemic bifurcation, $R_0=1$, must be $\G$-invariant, $I$-linear vaccination doesn't influence this threshold. This is in contrast to a constant vaccination rate, which is well known to reduce the reproduction number \cite{Driessche2017}.

Finally, the symmetry also covers the ``quasi-SIR limit'', defined by absence of constant vaccination and immunity waning. In this limit, we either have a pure SIS model or the model becomes $\G$-equivalent to a pure classic SIR model. Thus, the Hamiltonian formulation for these models carries over to the ``quasi-SIR'' case, see Appendix \ref{App_alpha=0}.

As an outlook let me mention, that the methods of this paper generalize to SI(R)S-type models with incomplete immunity, i.e. where also the $R$-compartment becomes susceptible. When including a social behavior term, the symmetry enlarges to $\G=\GS\times\GI\times\GX$, where now $\GS$ becomes non-abelian and is defined to be the sub-group of real $2\times 2$-matrices with positive determinant, acting on 
$(S,R)\in\RR^2$ and leaving $S+R$ invariant \cite{Nill_Symm1}. In combination with redundancy results for demographic parameters in \cite{Nill_Redundancy}, this covers a whole class of homogeneous SI(R)S-type models with time dependent total population, excess mortality and possibly also backward bifurcation 
\autocite{BusDries90, Mena-LorcaHeth, DerrickDriessche, Had_Cast, Had_Dries, KribsVel, LiMa2002, Vargas-De-Leon, dOnofrio_et_al_2018, Arino_et_al, AvramAdenane2022, AvramAdenane_et_al}.

\appendix

\section{The Korobeinikov-Wake SIRS model\label{App_KorW}}
This appendix shortly describes, within the present framework, the type of SIRS model introduced by Korobeinikov and Wake in \cite{KorobWake}. It will turn out that in a certain range of seemingly admissible parameters this model shows non-physical disease free and endemic equilibrium states satisfying $S^*\dfe>1$ and $S^*\en>1$.

The SIRS model  in \cite{KorobWake} introduces compartment dependent mortality rates $(\mu_S,\mu_I,\mu_R)$, keeping $\mu_S$ and $\mu_I$ time independent and postulating a time dependent $\mu_R(t)$, fine-tuned such that the total population $N$ stays constant. In this way, using the terminology of Fig. \ref{Fig_SIRS-Flow}, the dynamics of fractional variables becomes
\begin{align}
\dot{S} &=-(\be+\theta)SI+q_S\nu(S+I+R-p_II)-(\al_S+\mu_S)S+\ga_S I +\al_R R,
\\
\dot{I} &=\be SI-\tga I,\qquad \tga:=\ga_S+\ga_R+\mu_I-p_I\nu,
\\
\dot{R}&=-\dot{S}-\dot{I}.
\end{align}
Hence, also in this model demographic parameters become redundant, but the formulas \eqref{tilde_parameters} have to be replaced by

\begin{equation}
\begin{array}{rclrcl}
\tal_S &:=&\al_S+\mu_S-q_S\nu\,,\qquad\qquad &
\tga_S &:=&\ga_S+q_S(1-p_I)\nu\,,
\\
\tal_R &:=&\al_R+q_S\nu\,, &
\tga_R &:=&\tga-\tga_S=\ga_R+\mu_I-(q_R p_I+q_S)\nu\,.
\end{array}
\label{tilde_parameters_KW} 
\end{equation} 
In \cite{KorobWake} the authors considered their SIRS model without vaccination and with temporal immunity after recovery by putting 
$q_S=1$ 
and $\theta=q_R=\al_S=\ga_S=0$.
So, replacing $R=1-S-I$, this  model stays in the setting of Section \ref{Sec_dot-xy}, with $\tilde{\bfa}\in\Abio$, provided the birth rate $\nu$ is small enough,
$
\nu\leq\min\{\mu_S, \ga_R+\mu_I\}.
$
However, let's now consider the case $\al_R+\mu_S>0$ and
$$
0\leq\mu_S<\nu<(\ga_R+\mu_I)/p_I.
$$
Then $\tal_S<0$, $\tga>0$, $\tal_S+\tal_R>0$ and, by \eqref{f},
$$
\tilde{\rho}=(\tal_R-\tga_S)\be/\tga^2=(\al_R+p_I\nu)\be/\tga^2\geq 0.
$$
Hence, we are still in the setting of Section \ref{Sec_dot-xy}, but this time with $\tilde{\bfa}\in\AH\setminus\Abio$, due to the negative ``would-be'' vaccination rate $\tal_S$. In particular, irrespective of the value of $R_0$, under these conditions the disease free equilibrium in \eqref{dfe} will always be non-physical
$$
S^*\dfe=\frac{\tal_R}{\tal_R+\tal_S}=
\frac{\al_R+\nu}{\al_R+\mu_S}>1.
$$
By the same effect, putting $\tilde{r}_0:=\be/\tga$, the ''would-be vaccination reduced'' reproduction number actually satisfies 
$R_0=\tilde{r}_0S^*\dfe>\tilde{r}_0$, see \Eqref{f}. Hence, we could choose $0<\be<\tga$ such that
$$
\frac{\al_R+\mu_S}{\al_R+\nu}<\tilde{r}_0<1
$$
to get $R_0>1$ and therefore,
by Theorem \ref{Thm_SIRS} and Eq. \eqref{ee}, in this scenario we would have a globally stable endemic equilibrium as in \cite{KorobWake}, which, however, would also be non-physical, 
$$
S^*\en=1/\tilde{r}_0>1.
$$
Apparently, these scenarios have not been addressed by the authors in \cite{KorobWake}.

\section{The quasi-SIR Hamiltonian\label{App_alpha=0}}
\begin{figure}[H]
\includegraphics[width=0.5\textwidth]{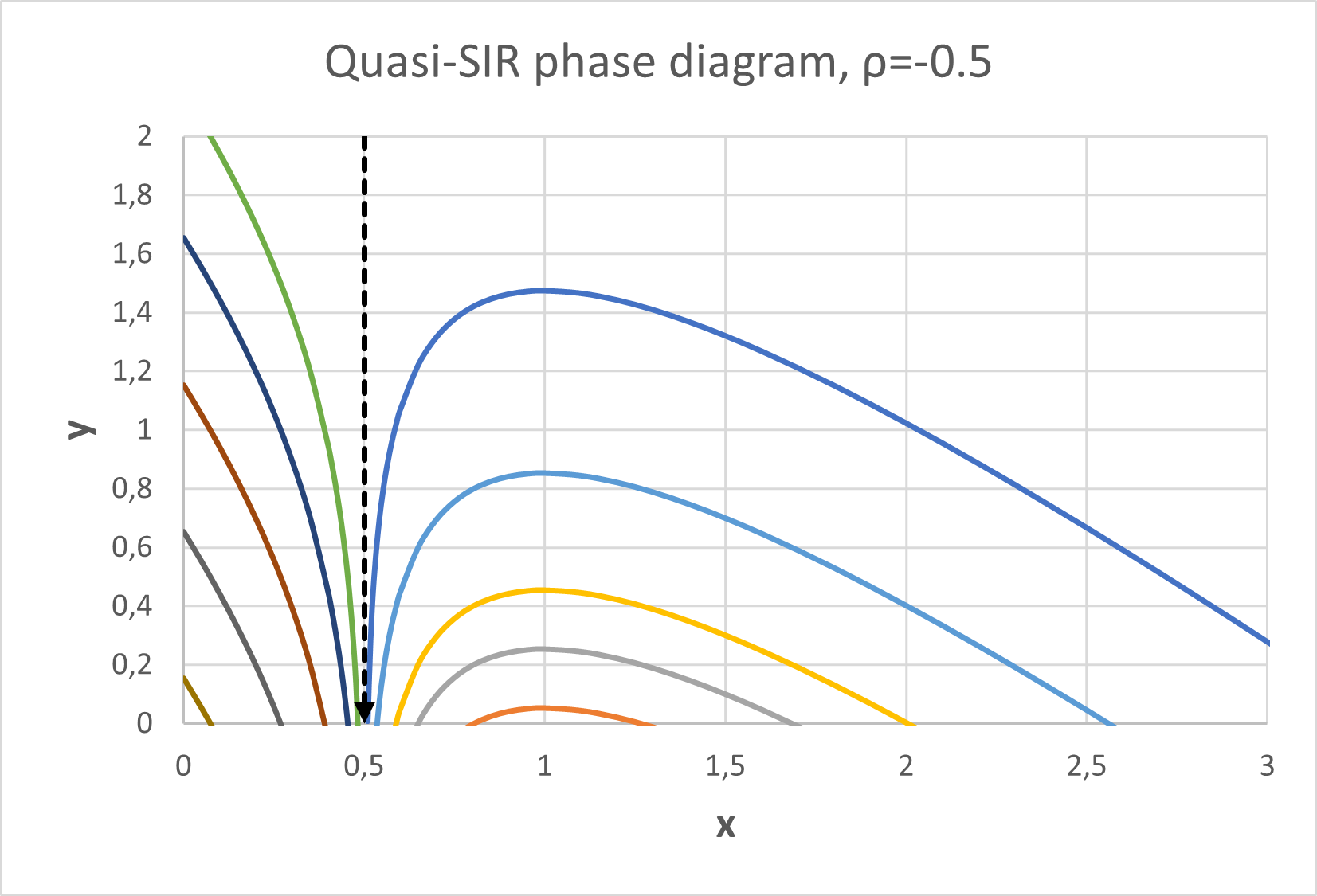}
	\caption{Phase diagram of the quasi-SIR model. $x(t)$ increases for $x_0<-\rho$ and decreases for $x_0>-\rho$. The vertical line corresponds to initial condition $x_0=-\rho$. }
	\label{Fig_quasi-SIR}
\end{figure}

To be self contained, this Appendix shortly studies the quasi-SIR limit, $\al_1=\al_2=0$ and $\theta_1\geq 0$, of the SI(R)S model \eqref{dot_bp}-\eqref{M+Lambda}.
This also leads to a proof of Theorem \ref{Thm_a=0}. Again, put $x=\be_1S$ and $y=\be_+y$ as in \eqref{f} to obtain
\begin{equation}
\begin{aligned}
\ga\inv\dot{x}	&=-(x+\rho)y,\qquad\rho=-\ga_1\be_1/\be_+\in[-1,0]
\\
\ga\inv\dot{y} &=(x-1)y.
\end{aligned}
\label{dot_xy_separating}
\end{equation}
This system factorizes, so we may choose
$\omega=(x+\rho)\inv y\inv$ as integrating factor, such that
$\omega\left[(x-1)ydx+(x+\rho)ydy\right]=dH$, with Hamiltonian  $H$ and symplectic form $\bom$ given by
\begin{align}
H		&= y+x-(\rho+1)\log|x+\rho|,
\\
\bom	&=-(\ga(x+\rho)y)\inv dx\wedge dy
\end{align}
\noindent
Note that $\rho=-1$ (i.e. $\ga_1=1$, $\ga_2=0$ and $\theta=0$) simplifies to the classic SIS model, $H=x+y$, and $\rho=0$ reproduces the classic SIR model Hamiltonian \cite{GuemNut}. For $0\geq\rho>-1$ we can apply the transformation \eqref{bar} to end up with the system 
\eqref{dot_txy_sirs} with $\bar{a}=0$, i.e. the classic SIR model in the variables $\bar{x}=(x+\rho)/(\rho+1)$ and $\bar{y}=y/(\rho+1)$. In all cases, phase space trajectories are lines of constant ``energy'', $H=\const$. Hence, they look like in the classic SIR model, extended to negative values, 
$\rho/(1+\rho)\leq\bar{x}<0$. So, we have a continuum of disease free equilibria, which are neutrally stable for $x<1\,(\LRA\bar{x}<1)$ and unstable for $x>1\,(\LRA\bar{x}>1)$. Also, at 
$x=-\rho\,(\LRA\bar{x}=0)$ we have an infinite energy barrier, which cannot be reached from either side, see Fig. \ref{Fig_quasi-SIR}. In fact, for initial condition $x_0=-\rho$ the explicit solution is given by the vertical line 
$(x(t),y(t))=(-\rho, y_0 e^{-\ga(\rho+1)t})$.

For initial conditions $0\leq x_0<-\rho<1$ and $y_0>0$ we have 
$\dot{x}(t)>0$ and 
$\dot{y}(t)<0$, for all $t\geq 0$, and $\lim_{t\to\infty}(x(t),y(t))=(x_\infty,0)$, where $x_0<x_\infty<-\rho<1$ solves 
$H(x_0,y_0)=H(x_\infty,0)$. In this case, no epidemic arises and 
$S(t)\uparrow S_\infty\equiv x_\infty/r_0$. 
This is due to the fact that in this region the recovery flow 
$\ga_S I$ exceeds the sum of infection $+$ vaccination flow 
$(\be+\theta)SI$.

If $x_0>-\rho$, solutions qualitatively behave like in the classical SIR model, i.e. $x(t)$ monotonically decreases with $\dot{y}>0$ for $x>1$, $\dot{y}<0$ for $x<1$ and $y=y_{\max}$ at $x=1$. Again, we have
$\lim_{t\to\infty}(x(t),y(t))=(x_\infty,0)$, where 
$H(x_0,y_0)=H(x_\infty,0)$, but this time $-\rho<x_\infty<1$. Note that $x_0>1$ necessarily requires $r_0>1$, i.e. no epidemic can arise for $r_0\leq 1$, as in the classic SIR model. Also, the inverse function $t(x)$ can be given explicitly, similarly as in \cite{HLM}.

\begin{lemma} 
Solutions of the dynamical system \eqref{dot_xy_separating} with initial conditions $x_0\neq-\rho$, $y_0>0$ and ``energy'' 
$E=H(x_0,y_0)$ satisfy
$$
\ga t =
-\int_{x_0}^{x(t)}\frac{dx}{(x+\rho)(E-x+(\rho+1)\log|x+\rho|)}\,,\qquad \sign(x(t)+\rho)=\sign(x_0+\rho).
$$
\end{lemma}
\begin{proof}
This follows immediately from 
$\ga\inv\dot{x}=-(x+\rho)y=-(x+\rho)(E-x+(\rho+1)\log|x+\rho|)$\,.
\end{proof}
Finally, the final size formula for $S_\infty$ as a function of $S_{-\infty}$ in Theorem \ref{Thm_a=0} is also obtained by ``energy'' conservation.

\bsn
{\bf Proof of Theorem \ref{Thm_a=0}:}
Part i) follows from the fact that in variables $(\bar{x},\bar{y})$ the system becomes a classic SIR model, where the initial conditions
$I_0>0$ and $S_0>\ga_1/(\be_1+\theta_1)$ translate to $\bar{y}_0>0$
and $\bar{x}_0>0$. Hence $\bar{x}_\infty<\bar{x}_0$ and 
$0<\bar{x}_\infty<1<\bar{x}_{-\infty}<\infty$.
To prove part ii) use
%
$$
\exp\left(\frac{H(S_\infty,0)}{\rho+1}\right)=
\exp\left(\frac{H(S_{-\infty},0)}{\rho+1}\right) \quad\Longrightarrow\quad
\bar{x}_\infty e^{-\bar{x}_\infty}=
\bar{x}_{-\infty} e^{-\bar{x}_{-\infty}},
$$
\Eqref{final_size} follows from $\be_1\equiv r_0$ and $\bar{x}=(r_0S+\rho)/(\rho+1)$.
%
To prove part iii) use
$$
\int_{-\infty}^\infty\dot{I}dt=0\quad\Longrightarrow\quad
\be\int_{-\infty}^\infty SIdt=\ga\int_{-\infty}^\infty Idt.
$$
Hence we get
$$
\Del R=\int_{-\infty}^\infty \dot{R}dt=
\ga\left(\theta_1 \int_{-\infty}^\infty SIdt + 
\ga_2\int_{-\infty}^\infty Idt\right)=
\ga\theta_1 \int_{-\infty}^\infty SIdt + 
\ga_2\be\int_{-\infty}^\infty SIdt\,,
$$
where the first term gives the fraction of vaccinated people. Part iii) follows from $r_0=\be/\ga$.
\qed

\section{The SIS model with vaccination\label{App_SIS+vacc}}
The transformation \eqref{bar} from the SI(R)S model to Hethcote's model becomes ill-defined for $\rho=-1$. 
Epidemiologically, the model with parameters $\bfa\in\Abio\cap\{\rho=-1\}$ is uninteresting and near to trivial. It implies 
$\al_2=\ga_2=\theta_1=0$, whence also $R_0=0$, see Remark \ref{Rem_rho=-1}. This is a pure SIS model furnished with a constant vaccination rate from $S$ to $R$ and permanent immunity in $R$. So, eventually all people are vaccinated and this model only shows the trivial equilibrium 
$(S^*,I^*,R^*)=(0,0,1)$. Global stability in $\Tph$ follows from absence of periodic solutions (use $I\inv$ as a Dulac function as in \cite{Hethcote1976, Hethcote1989}) and the fact that $\Tph$ is forward invariant.

\printbibliography

\end{document}